\documentclass[twocolumn]{aastex62}

\submitjournal{ApJ}

\graphicspath{{./}{figures/}}

\received{January 20, 2020}

\submitjournal{ApJ}

\shortauthors{Bennet et al.}

\begin{document}

\title{{\it Hubble Space Telescope} Imaging of Isolated Local Volume Dwarfs GALFA-Dw3 and Dw4}

\correspondingauthor{Paul Bennet}
\email{pbennet@stsci.edu}

\author[0000-0001-8354-7279]{P. Bennet}
\affiliation{Physics \& Astronomy Department, Texas Tech University, Box 41051, Lubbock, TX 79409-1051, USA}
\affiliation{Space Telescope Science Institute, 3700 San Martin Drive, Baltimore, MD 21218, USA}
\author[0000-0003-4102-380X]{D. J. Sand}
\affiliation{Department of Astronomy/Steward Observatory, The University of Arizona, 933 North Cherry Avenue, Rm. N204, Tucson, AZ 85721-0065, USA}
\author[0000-0002-1763-4128]{D. Crnojevi\'{c}}
\affiliation{University of Tampa,  Department of Chemistry, Biochemistry, and Physics, 401 West Kennedy Boulevard, Tampa, FL 33606, USA}
\author{D. R. Weisz}
\affiliation{University of California, Berkeley, Department of Astronomy, 501 Campbell Hall \# 3411, Berkeley, CA 94720-3411,USA}
\author[0000-0003-2352-3202]{N. Caldwell}
\affiliation{Harvard-Smithsonian Center for Astrophysics, 60 Garden Street, Cambridge, MA 02138, USA}
\author[0000-0001-8867-4234]{P. Guhathakurta}
\affiliation{UCO/Lick Observatory, University of California, Santa Cruz, 1156 High Street, Santa Cruz, CA 95064, USA}
\author[0000-0002-8722-9806]{J. R. Hargis}
\affiliation{Space Telescope Science Institute, 3800 San Martin Drive, Baltimore, MD, 21208, USA}
\author[0000-0001-8855-3635]{A. Karunakaran}
\affiliation{Department of Physics, Engineering Physics and Astronomy, Queen’s University, Kingston, ON K7L 3N6, Canada}
\author[0000-0001-9649-4815]{B. Mutlu-Pakdil}
\affiliation{Kavli Institute for Cosmological Physics, University of Chicago, Chicago, IL 60637, USA}
\affiliation{Department of Astronomy and Astrophysics, University of Chicago, Chicago IL 60637, USA}
\author[0000-0002-7157-500X]{E. Olszewski}
\affiliation{Department of Astronomy/Steward Observatory, The University of Arizona, 933 North Cherry Avenue, Rm. N204, Tucson, AZ 85721-0065, USA}
\author[0000-0001-8483-603X]{J. J. Salzer}
\affiliation{Department of Astronomy, Indiana University, 727 East Third Street, Bloomington, IN 47405, USA}
\author{A. C. Seth}
\affiliation{Department of Physics and Astronomy, University of Utah, 115 South 1400 East, Salt Lake City, Utah 84112, USA}
\author{J. D. Simon}
\affiliation{Observatories of the Carnegie Institution for Science, Pasadena, California 91101, USA}
\author[0000-0002-0956-7949]{K. Spekkens}
\affiliation{Department of Physics and Space Science, Royal Military College of Canada P.O. Box 17000, Station Forces Kingston, ON K7K 7B4, Canada}
\affiliation{Department of Physics, Engineering Physics and Astronomy, Queen’s University, Kingston, ON K7L 3N6, Canada}
\author[0000-0001-6106-5172]{D. P. Stark}
\affiliation{Department of Astronomy/Steward Observatory, The University of Arizona, 933 North Cherry Avenue, Rm. N204, Tucson, AZ 85721-0065, USA}
\author{J. Strader}
\affiliation{Center for Data Intensive and Time Domain Astronomy, Department of Physics and Astronomy, Michigan State University, East Lansing, MI 48824, USA}
\author{E. J. Tollerud}
\affiliation{Space Telescope Science Institute, 3700 San Martin Drive, Baltimore, MD 21218, USA}
\author{E. Toloba}
\affiliation{Department of Physics, University of the Pacific, 3601 Pacific Avenue, Stockton, CA 95211, USA}
\author{B. Willman}
\affiliation{LSST and Steward Observatory, 933 North Cherry Avenue, Tucson, AZ 85721, USA}

\begin{abstract}
We present observations of the dwarf galaxies GALFA Dw3 and GALFA Dw4 with the Advanced Camera for Surveys (ACS) on the Hubble Space Telescope (HST). These galaxies were initially discovered as optical counterparts to compact HI clouds in the GALFA survey. Both objects resolve into stellar populations which display an old red giant branch, younger helium burning, and massive main sequence stars. We use the tip of the red giant branch method to determine the distance to each galaxy, finding distances of 7.61$_{-0.29}^{+0.28}$ Mpc and 3.10$_{-0.17}^{+0.16}$ Mpc, respectively. With these distances we show that both galaxies are extremely isolated, with no other confirmed objects within $\sim$1.5 Mpc of either dwarf. GALFA Dw4 is also found to be unusually compact for a galaxy of its luminosity.
GALFA Dw3 \& Dw4 contain HII regions with young star clusters and an overall irregular morphology; they show evidence of ongoing star formation through both ultraviolet and H$\alpha$ observations and are therefore classified as dwarf irregulars (dIrrs). 
The star formation histories of these two dwarfs show distinct differences: Dw3 shows signs of a recently ceased episode of active star formation across the entire dwarf, while Dw4 shows some evidence for current star formation in spatially limited HII regions.  
Compact HI sources offer a promising method for identifying isolated field dwarfs in the Local Volume, including GALFA Dw3 \& Dw4, with the potential to shed light on the driving mechanisms of dwarf galaxy formation and evolution.  
\end{abstract}


\section{Introduction} \label{sec:intro}

The Lambda Cold Dark Matter model for structure formation has been very successful at reproducing observations of large scale structures; however challenges emerge at sub-galactic scales \citep[for a recent review, see][and the references therein]{Bullock17}. Some of these challenges  
can be examined by switching focus from dwarf galaxies in nearby groups \citep{McConnachie18,Crnojevic19,Bennet19, Bennet20,Carlsten20,Mao20} 
to isolated field galaxies within the Local Volume \citep{sand15, McQuinn15b,tollerud16}. 

Examining these isolated, gas rich, dwarf galaxies is critical to our understanding of dwarf galaxy formation and testing dark matter theories. They are the faintest/least massive galaxies we know of that have never interacted with a massive galaxy halo, and thus have never felt the effects of tidal/ram pressure stripping \citep{spekkens14,Wetzel15}. They are a more controlled experiment for understanding other mechanisms which drive the star formation history (SFH) and metallicity of a dwarf galaxy, for instance supernova-driven winds, or infall of pristine gas from the local environment \citep[][]{mcquinn13}. 
By characterizing their resolved stellar populations, it becomes possible both to obtain the present-day structural parameters for these galaxies and to characterize their SFHs, providing constraints on their pasts \citep{mcquinn15,tollerud16,McQuinn20}. 
Additionally, these gas rich galaxies potentially trace the full dwarf galaxy population at the outskirts of the Local Group and other similar low-density environments, a regime where the numbers and properties of these dwarfs are just starting to be compared directly with numerical simulations \citep[][]{Tikhonov09,Garrison14,Garrison19,Tollerud18}. 

In this work, we will examine the isolated Local Volume dwarf galaxies GALFA Dw3 and Dw4. These objects were discovered as part of an archival search for optical counterparts to HI clouds \citep[][]{Giovanelli10} discovered in the ALFALFA \citep{adams13} and GALFA \citep{saul12} surveys by \citet{sand15}, and were both confirmed to have H$\alpha$ emission at a velocity consistent with the HI detection. 
The key properties of GALFA Dw3 and Dw4 are listed in Table \ref{tab:prop}.

An outline of the paper follows. In Section \ref{sec:data}, we describe the {\it HST} photometry and artificial star tests (ASTs), as well as supplemental observations of the dwarfs. In Section \ref{sec:TRGB}, we derive distances to GALFA Dw3 and Dw4 via the Tip of the Red Giant Branch (TRGB) method. In Section \ref{sec:structure}, we examine the observational properties of the dwarfs in the {\it HST} imaging and derive their physical properties. In Section \ref{sec:star_form}, we discuss the star formation histories based on their {\it HST} color-magnitude diagrams (CMDs), as well as supplemental H$\alpha$ and ultraviolet (UV) images. In Section \ref{sec:disc}, we discuss the environment of the dwarfs and potential analogs within the Local Volume. Finally we summarize and conclude in Section \ref{sec:conclusion}.

\section{Data Overview} \label{sec:data}

\subsection{Hubble Space Telescope Observations}

\begin{figure*}
 \begin{center}
 \includegraphics[width=10cm]{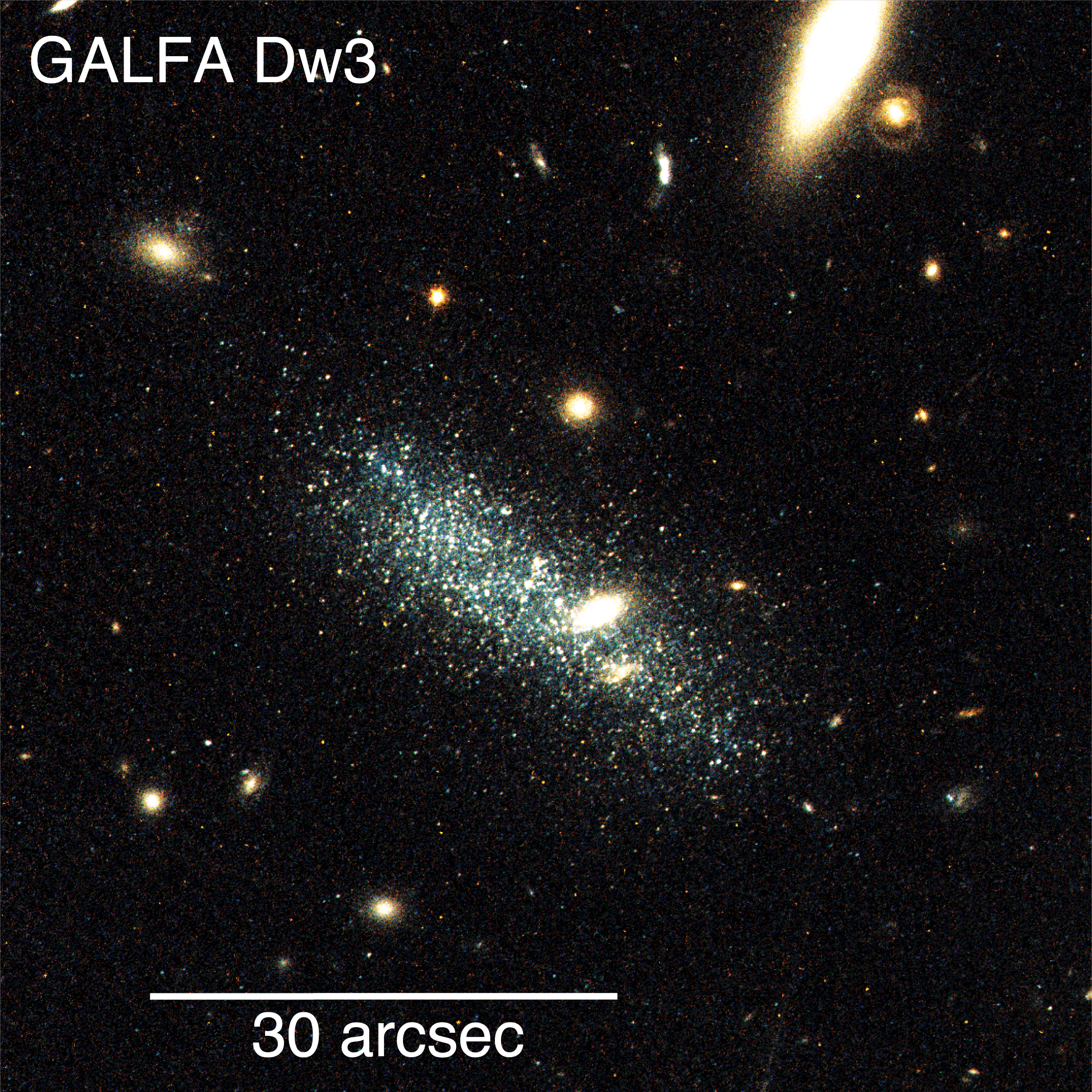}
 \includegraphics[width=10cm]{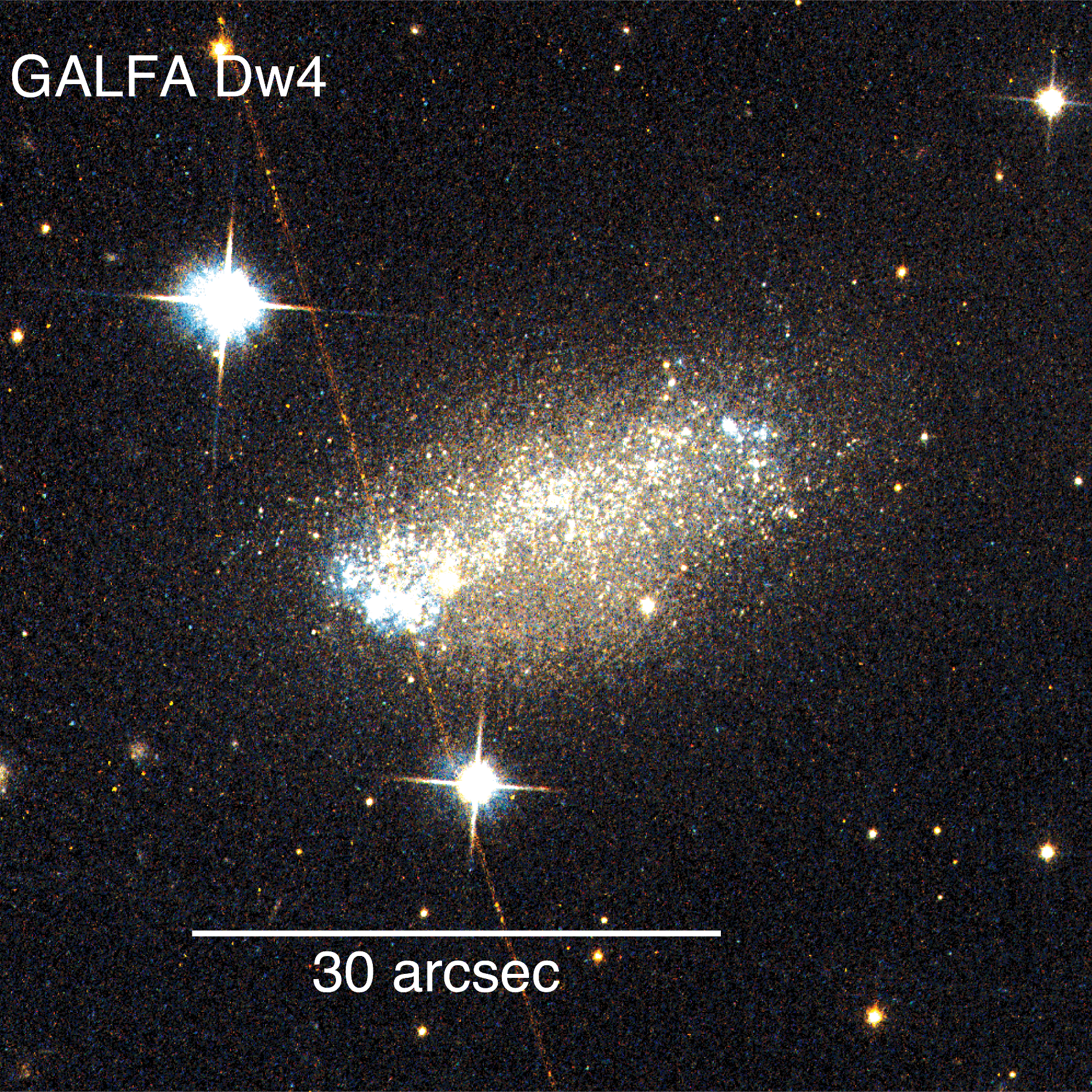}
 \caption{Color composite of F606W/F814W {\it HST} ACS imaging of the dwarf galaxies GALFA Dw3 (upper panel) and GALFA Dw4 (lower panel). The bright objects in the SW of Dw3 are background galaxies. 
 Images are 1.2'x1.2'. North is up, east is left. \label{fig:gal_col}}
 \end{center}
\end{figure*}

The {\it HST} observations of GALFA Dw3 \& Dw4 were taken as part of program GO-14676 (Cycle 24, PI Sand). Both Dw3 \& Dw4 were observed for a single orbit with the  Advanced Camera for Surveys (ACS)/Wide Field Camera (WFC), using the F606W and F814W filters. We did not dither to fill in the WFC CCD chip gap, as each dwarf easily fit into one chip. The total exposure time was 1062 s for each filter on both Dw3 \& Dw4. Color composites of these images are shown in Figure~\ref{fig:gal_col}.

We perform PSF-fitting photometry on the provided {\it .flt} images using the DOLPHOT v2.0 photometric package (with the ACS module), a modified version of HSTphot \citep{dolphin00}. For this work we use the suggested input parameters from the DOLPHOT/ACS User's Guide\footnote{\url{http://americano.dolphinsim.com/dolphot/dolphotACS.pdf}}, including corrections for charge transfer efficiency losses and default aperture corrections based around a 0.5'' aperture. Quality cuts are then applied using the following criteria: the derived photometric errors must be $\leq$0.3 mag in both bands, the sum of the crowding parameter in both bands is $\leq$1 and the square of the sum of the sharpness parameter in both bands is $\leq$0.075. Detailed descriptions of these parameters can be found in \cite{dolphin00}. 
For this analysis, we correct these extracted magnitudes for foreground extinction and reddening using the \cite{Schlafly11} calibration of the \cite{schlegel98} dust maps (we note that GALFA Dw4 suffers from significant extinction due to its proximity to the plane of the Galaxy, E(B-V)=0.531 mag).  

We estimate photometric uncertainties using ASTs in a CMD region covering the full range of observed stars, from blue Main Sequence (MS) features to regions redward of the red giant branch (RGB). The fake stars have a similar color-magnitude distribution to that of the observed sources, except for a deeper extension at faint magnitudes (down to $\sim$2 mag fainter than the faintest real recovered stars), so as to take into account those faint objects that are upscattered in the observed CMD due to noise. The AST photometry is derived in exactly the same way as for the real data, and the same quality cuts and calibration are applied. 

The resulting CMDs can be seen in Figure~\ref{fig:gal_CMD}. The completeness and uncertainties for Dw4 appear to be worse than that of Dw3, but this is solely because of the higher extinction associated with Dw4; they are identical in uncorrected apparent magnitude space. 

We assessed the crowding for each field. Visual inspection of GALFA Dw3 showed clearly separated point sources throughout the main body of the dwarf. 
GALFA Dw4 required more careful examination, with possible crowding in the blue knots in the southeast and northwest ends of the dwarf (see Figure \ref{fig:gal_col}). Examination of the potentially crowded regions showed similar completeness levels to those found in the rest of the dwarf when using standard photometry, and visual inspection showed no obviously missed point sources in the region in question. 
We also made standard changes to the photometry recommended for crowded regions, namely setting the parameter FitSky=3 (for more details please see the DOLPHOT's User Guide). 
This crowded photometry was then compared to the standard photometry in the affected region with no significant difference between the two: we conclude that the use of crowded photometry parameters was unnecessary and that standard parameter photometry was as effective in all regions of GALFA Dw4. However, 
some of the stars from Dw4 may not be recovered successfully in either the standard or crowded photometry, and this will be further discussed in \S\ref{subsec:Dw4}. 

\subsection{Other Observations}

\begin{figure*}
 \begin{center}
 \includegraphics[width=5cm]{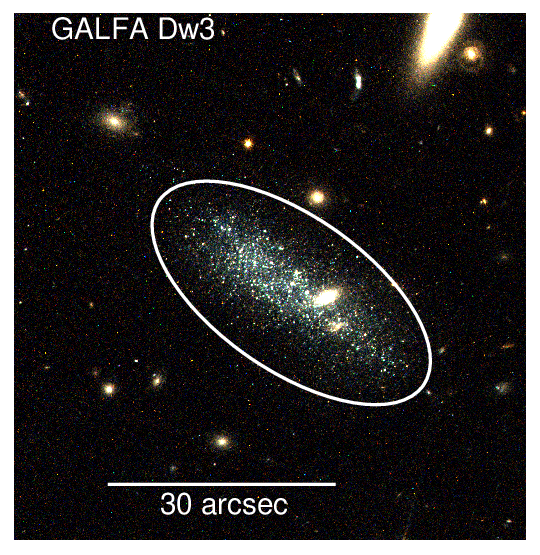}
 \includegraphics[width=5cm]{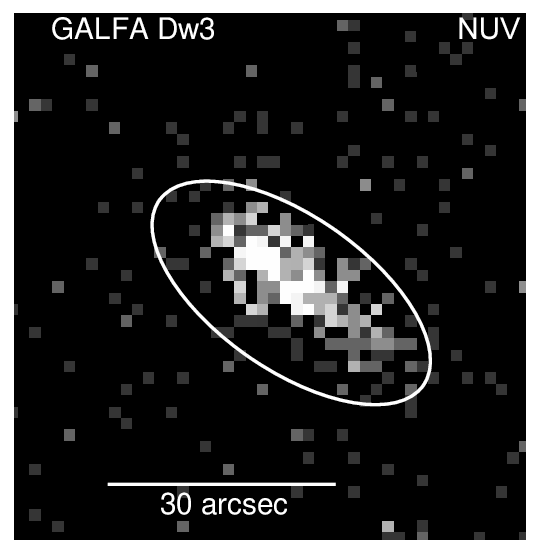}
 \includegraphics[width=5cm]{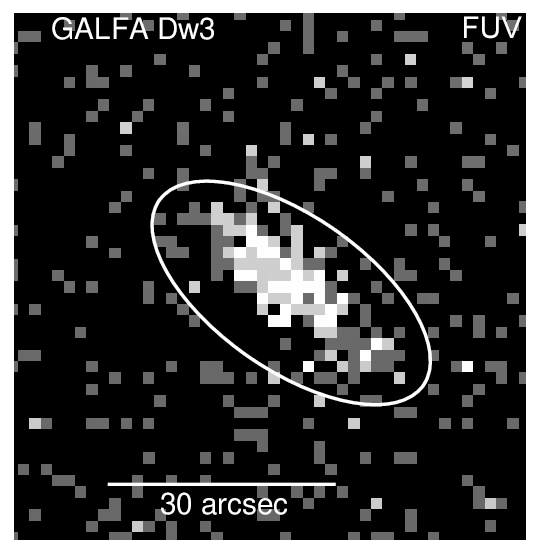}
 \caption{The UV images of GALFA Dw3 from the GALEX All Sky Imaging Survey (AIS) alongside optical images from {\it HST} for illustrative purposes, see Figure \ref{fig:gal_col}. This clearly shows the elevated UV emission from Dw3. North is up, east is left. Each image is 1.1'x1.1'. The ellipses in this plot are illustrative. Left: HST Optical, Center: GALEX NUV, Right: GALEX FUV. \label{fig:gal_3_UV}}
 \end{center}
\end{figure*}

Data from the Galaxy Evolution Explorer (GALEX; \citealt{martin05}) were also used to check for UV emission from GALFA Dw3, as this can be a strong indicator of recent star formation. Indeed, GALFA Dw3 shows substantial FUV and NUV emission, which we report alongside the {\it HST} data in Figure~\ref{fig:gal_3_UV}. These data were part of the All-Sky Imaging Survey; see \cite{Morrissey07} for details. GALFA Dw4 is outside the GALEX footprint and therefore no conclusions can be drawn about its recent star formation with this dataset. We thus used UV images from the Neil Gehrels $Swift$ Observatory \citep{Gehrels04} and the Ultraviolet/Optical Telescope \citep[UVOT; ][]{swift_uvot}, which were taken as part of proposal 1417202 (P.I. L. Hagen) in all 3 available UV filters (UVW1, UVM2, UVW2). There is no UV emission detected in these data, likely due to the high levels of extinction along the line of sight to Dw4.  

\begin{figure*}
 \begin{center}
 \includegraphics[width=7cm]{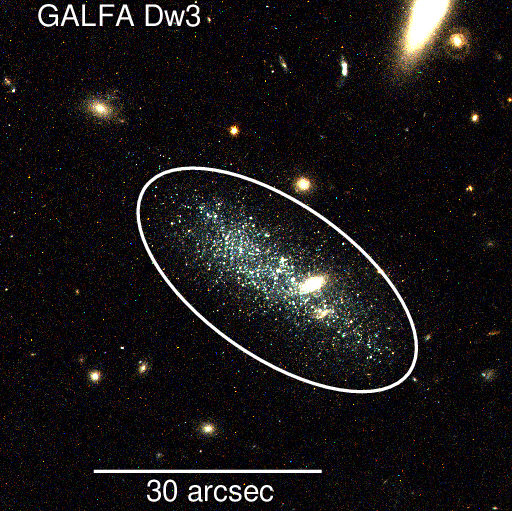}
 \includegraphics[width=7cm]{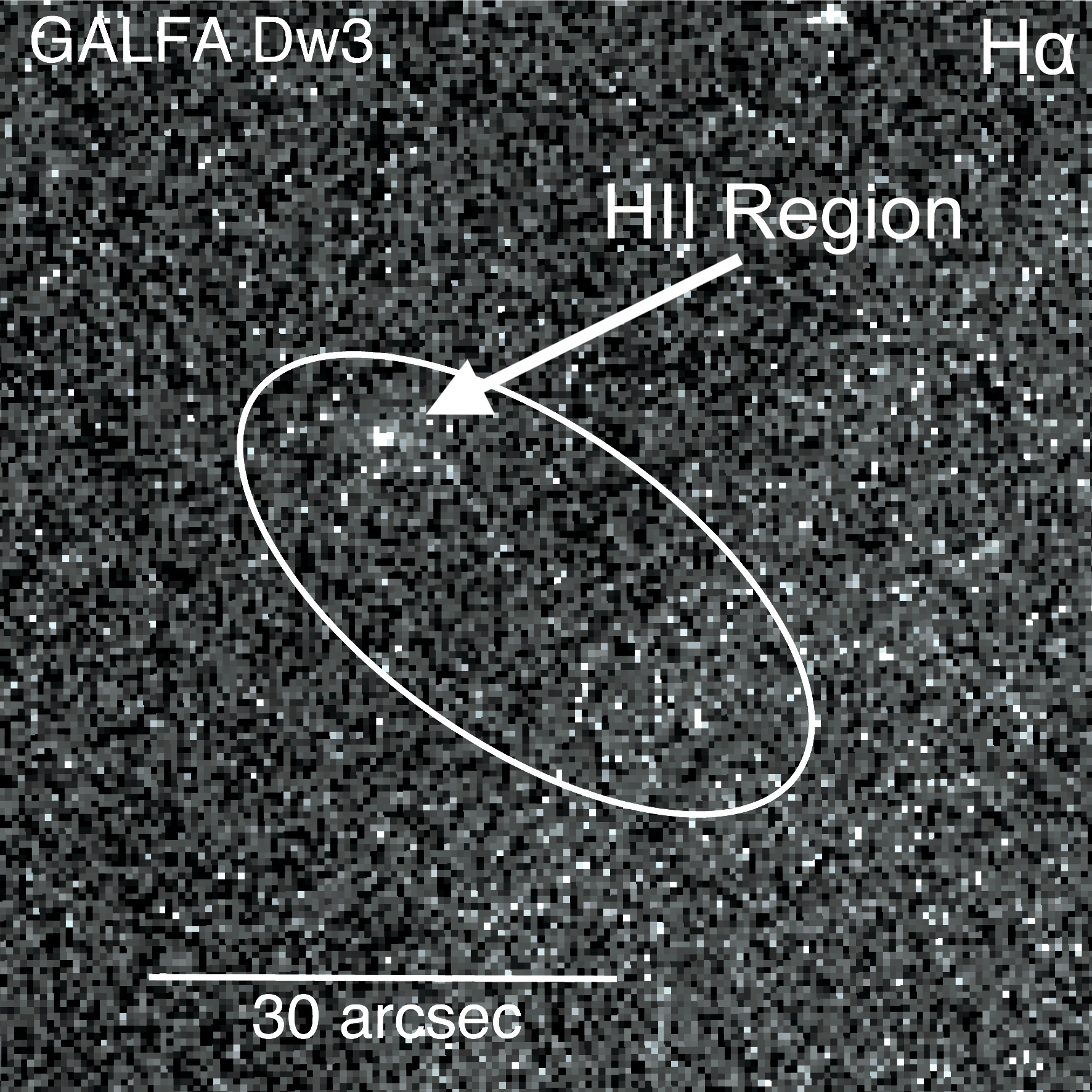}
 \includegraphics[width=7cm]{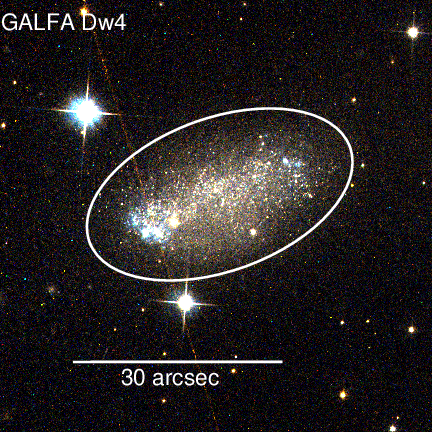}
 \includegraphics[width=7cm]{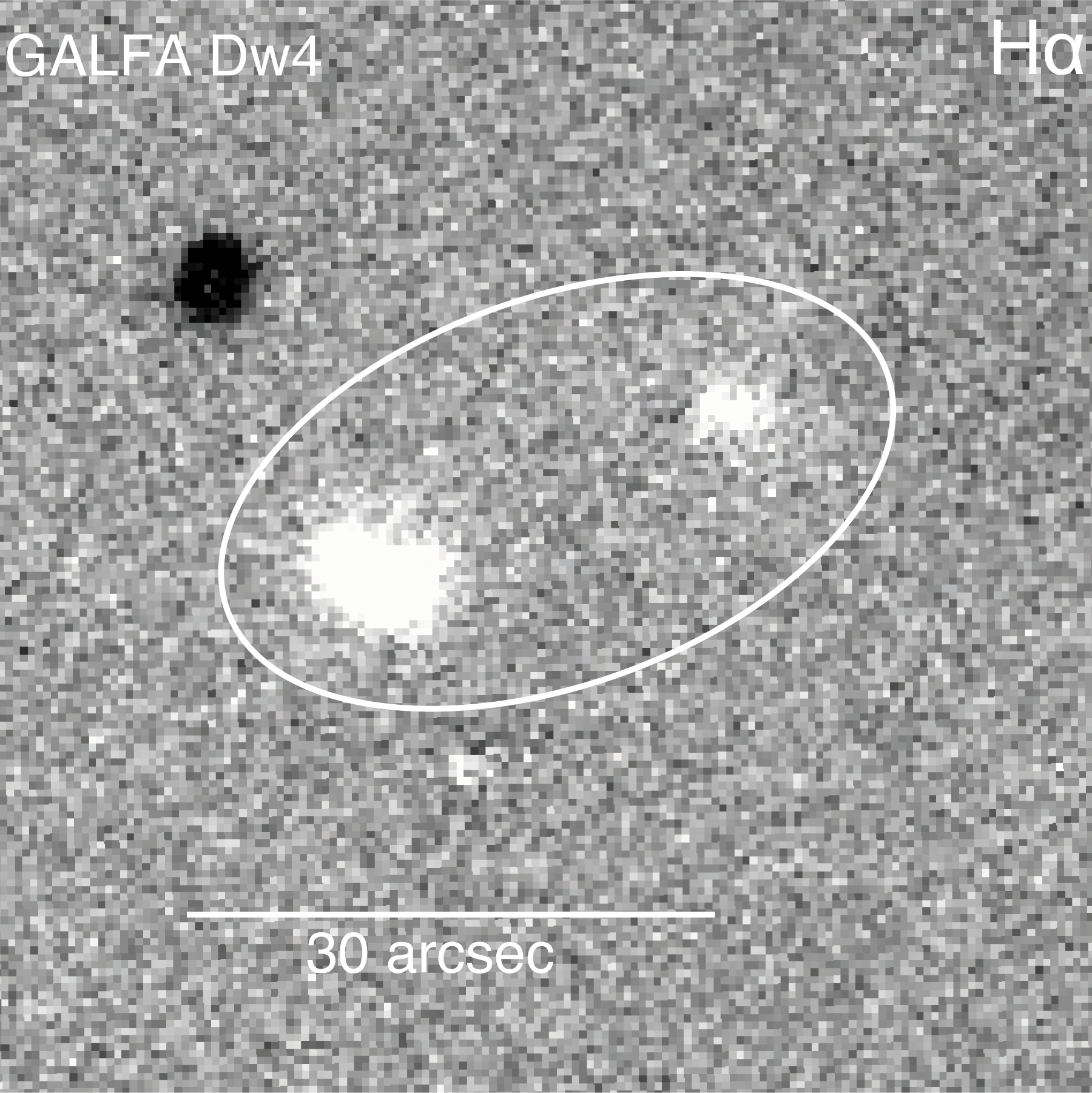}
 \caption{The H$\alpha$ narrow band images (see \S\ref{sec:data}) 
 of GALFA Dw3 and Dw4 minus the continuum emission (right column), alongside optical images from {\it HST} for illustrative purposes (left column). We point out the elevated H$\alpha$ emission from the northeast corner of Dw3. GALFA Dw4 shows more H$\alpha$ emission within two clear regions, one at the southeast end of the dwarf and the other at the northwest end. These regions match with the blue regions seen in the {\it HST} imaging. North is up, east is left. Each image is 1.1'x1.1'. The ellipses in this plot are illustrative. 
 \label{fig:gal_Ha}}
 \end{center}
\end{figure*}

Supplemental H$\alpha$ narrow band imaging of GALFA Dw3 \& Dw4 were obtained by our group with the WIYN 0.9-m telescope and the Half Degree Imager on 21 July 2017 (UT). These images are used to trace HII regions with active star formation within the last $\sim$10 Myrs \citep{Calzetti13} and can be seen in Figure \ref{fig:gal_Ha}. 

\section{Tip of the Red Giant Branch Distances} \label{sec:TRGB}

To determine distances to these resolved dwarf galaxies, we make use of the TRGB technique \citep[e.g.,][]{dacosta90,lee93,Makarov06,Rizzi07,Freedman20}. The peak luminosity of the RGB is a standard candle in the red bands, because it is driven by core helium ignition and so it provides a useful distance estimate for galaxies with an old stellar component which are close enough that the RGB stars can be resolved. To determine TRGB magnitudes, we adopt the methodology described in \cite{Crnojevic19}. Briefly, the photometry is first corrected to account for the color dependence of the TRGB \citep{jang17}; we also consider only RGB stars with colors in the range $0.85<(F606W-F814W)_0<1.35$, so as to exclude possible contamination from young red supergiant stars. The luminosity function for RGB stars is then computed (note that the field, background+foreground, contamination as derived from a dwarf-free region of the ACS field-of-view is not significant for the range of colors and magnitudes considered here), and a model luminosity function (convolved with the appropriate photometric uncertainty, bias and incompleteness function as derived from our ASTs) is fit to it with a non-linear least squares method. 

\begin{figure*}
 \begin{center}
 \includegraphics[width=8cm]{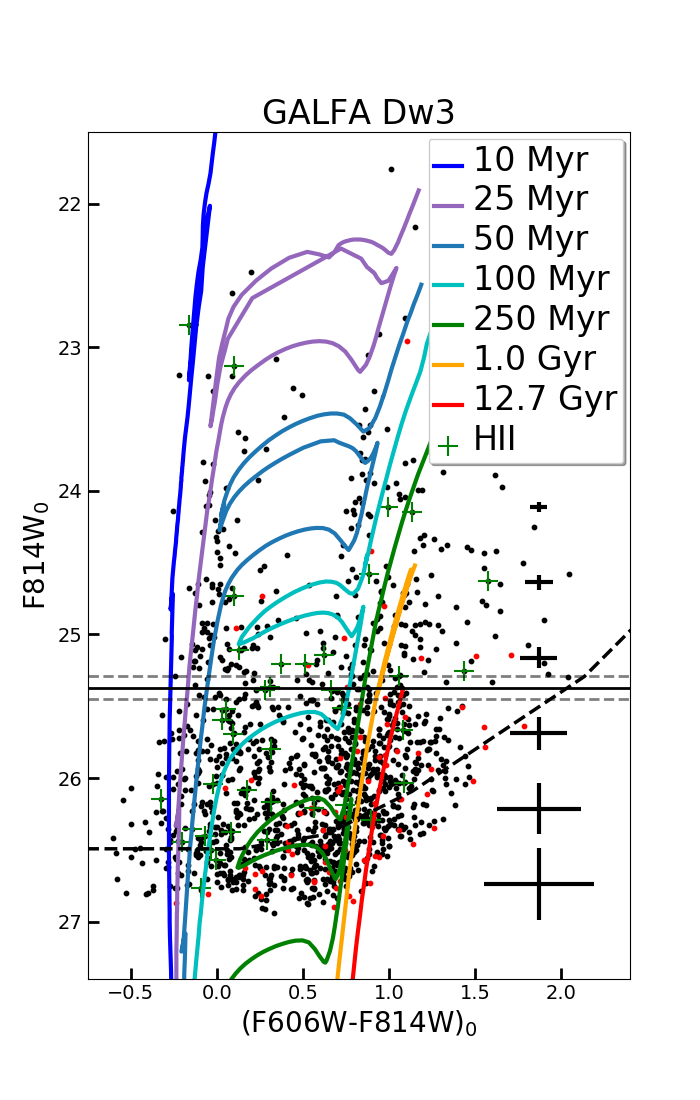}
 \includegraphics[width=8cm]{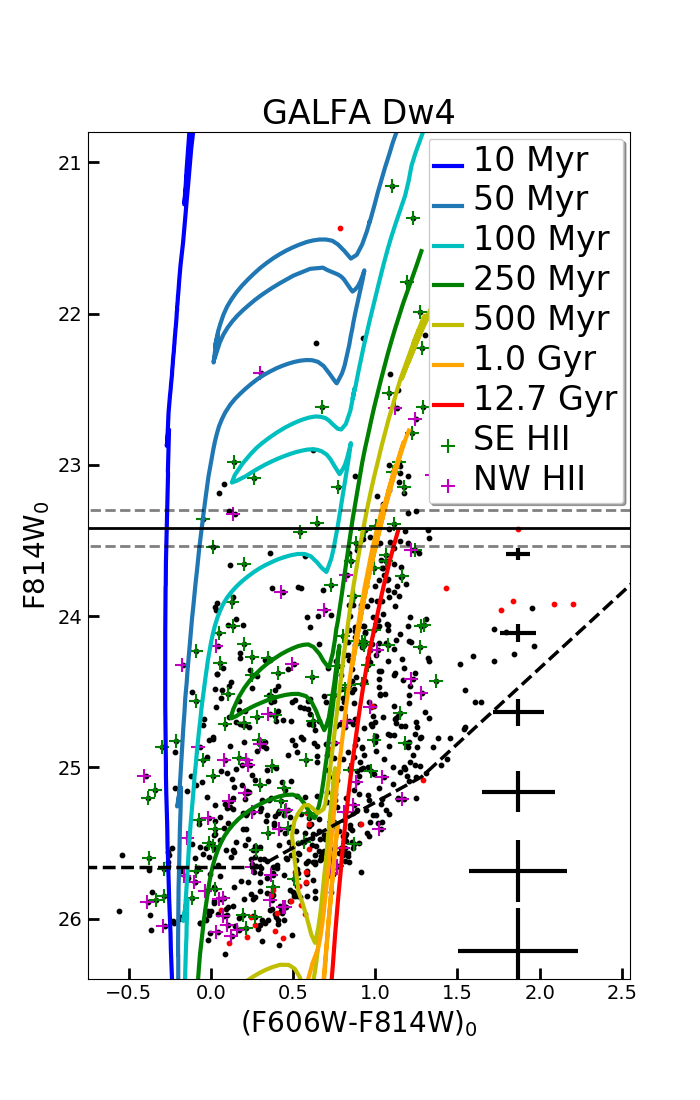}
 \caption{F606W/F814W CMD for the dwarf galaxies GALFA Dw3 (left panel) and GALFA Dw4 (right panel). Magnitudes are corrected for foreground extinction (see \S\ref{sec:data}). 
Only point sources are shown (i.e., those sources with a DOLPHOT object type=1 or 2). Black dots are stars within the dwarfs, red dots are stars from an equal-area control field. In the left panel, the green crosses indicate those stars associated with the spatial position of the HII region in Dw3, see \S\ref{subsec:Dw3}. In the right panel, the green crosses indicate those stars associated with the spatial position of the southeast HII region and the magenta crosses those associated with the northwest HII region, see \S\ref{subsec:Dw4}. The black horizontal line indicates the best fit for the TRGB, and the dashed gray lines represent the 1$\sigma$ uncertainty. We display several Padova isochrones \citep{Bressan12}, shown as solid lines of varying color, each line representing a stellar population of fixed age, shown in the legend of each panel. The red isochrone (RGB stars) is plotted at [Fe/H]=$-$1.6 for both dwarfs, while all other isochrones are at [Fe/H]=$-$1.0. Finally, the 50$\%$ completeness limit (black dashed line) and the photometric uncertainties are reported.
\label{fig:gal_CMD}}
 \end{center}
\end{figure*}

Using the {\it HST} data, we find a TRGB magnitudes of 25.37$\pm$0.08 mag and 23.42$\pm$0.12 mag for GALFA Dw3 and Dw4, this correspond to distance moduli of 29.41$\pm$0.08 and 27.46$\pm$0.12 mag, which translate to distances of 7.61$_{-0.29}^{+0.28}$ Mpc and 3.10$_{-0.17}^{+0.16}$ Mpc, respectively. 
We mark the position of the TRGB and its uncertainty in Figure~\ref{fig:gal_CMD}, and tabulate our results in Table~\ref{tab:prop}.

\cite{Anand19} used the same dataset presented here for GALFA Dw4 to study the peculiar velocities of galaxies at the edge of the Local Group, and reported a TRGB distance of 2.97$\pm$0.37 Mpc, which is consistent with the distance reported here. 

\section{Structural Parameters} \label{sec:structure}

Utilizing the {\it HST} imaging, we revisit the structural properties of these dwarf galaxies, previously reported in \cite{sand15}. To constrain the structural parameters, we use the maximum-likelihood technique of \citet{Martin08} using the implementation of \citet{Sand09,Sand12}. First, we select the stars consistent with the RGB as seen in Figure \ref{fig:gal_CMD}. We fit a standard exponential profile plus constant background to the data, with the following free parameters: the central position (RA$_0$, DEC$_0$), position angle, ellipticity, half-light radius ($r_h$) and background surface density. 
Uncertainties on structural parameters are determined by bootstrap resampling the data 1000 times, from which 68\% confidence limits are calculated.  
The resulting structural parameters are summarized in Table \ref{tab:prop}.
 
\begin{figure*}
 \begin{center}
 \includegraphics[width=8cm]{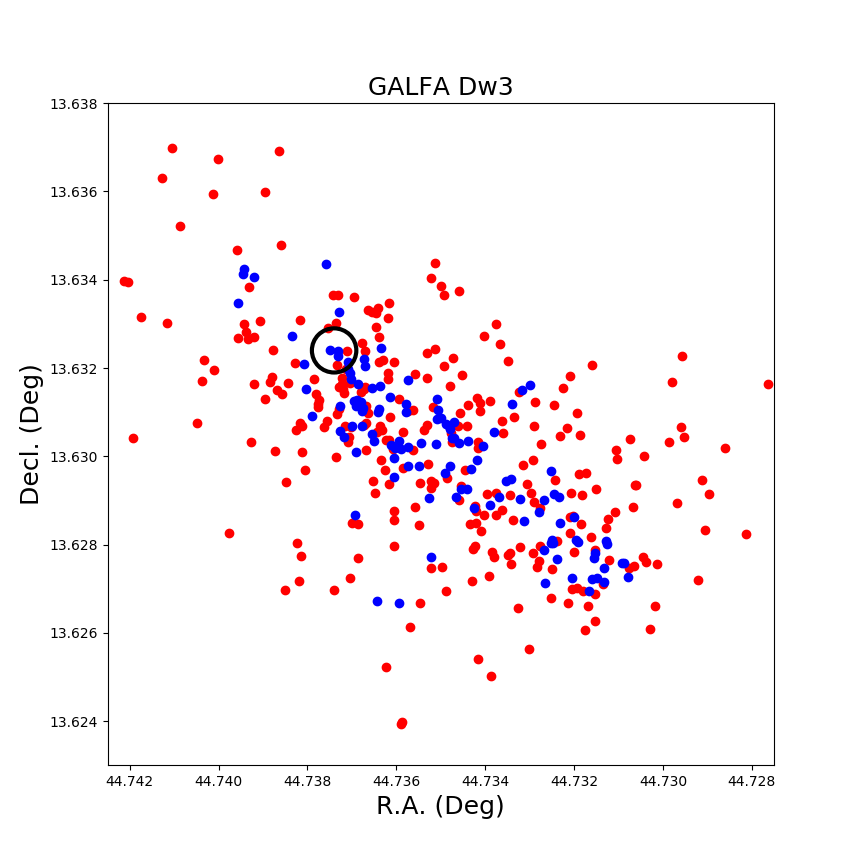}
 \includegraphics[width=8cm]{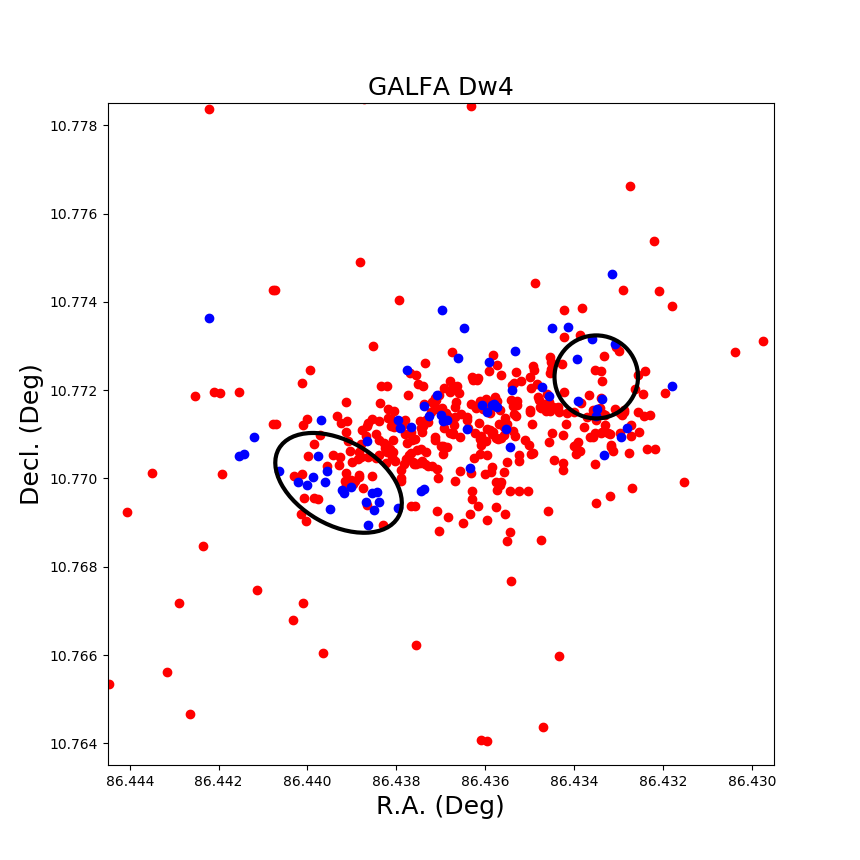}
 \caption{ 
 Spatial distribution of point sources consistent with stellar populations in GALFA Dw3 and Dw4. Point sources consistent with RGB stars are shown in red; these are selected via matching to the RGB isochrones seen in Figure \ref{fig:gal_CMD}. The blue points are those point sources consistent with a color of (F606W$_0$-F814W$_0$)$<$0.1, which are consistent with MS and blue helium burning stars. Only stars brighter than our 50\% completeness limits are plotted. 
 The approximate position and size of the HII regions in both dwarfs are shown by black outlines. 
 The blue stars in Dw3 have a higher ellipticity than the RGB populations, but are generally spread throughout the dwarf. In Dw4 there is a concentration of blue stars around the HII region to the southeast, along with several associated with the HII region to the northwest, but few in the main body of the dwarf. 
 Panels are 0.9\arcmin squares. North is up, East is left. 
 \label{fig:star_map}}
 \end{center}
\end{figure*}

Note that while the derived parameters describe the older stellar populations in our targets, both Dw3 and Dw4 host  young populations that are highly irregular in appearance and are concentrated in the HII regions in the case of Dw4 (see Figure \ref{fig:star_map}).

We derive the absolute magnitude of the dwarfs via direct aperture photometry using an elliptical aperture with semi-major axis equal to the half-light radius. We estimate the flux within this aperture (after background correction), and multiply by a factor of two to account for the total flux of the dwarf, and then convert to a magnitude. After applying our measured distance modulus and correction for galactic extinction, we find M$_{V}=-12.8\pm0.3$ and $-11.8\pm0.3$ for Dw3 and Dw4, respectively.  Our results are consistent with the properties reported in \cite{sand15} within the uncertainties. 
We then estimate the present day stellar mass from the V band luminosity combined with the V-I color using the mass to light ratio formalism from \cite{Bell01}: 

\begin{equation}
\log(M/L)_{V} = a_{V} + b_{V} \cdot (V-I)
\end{equation}

\noindent where a$_{V}$$=$$-$1.476 and b$_{V}$$=$1.747 with an assumed solar luminosity of M$_{V}$$=$4.77. This produces masses of 2.1$\times$10$^6$ M$_\odot$ and 2.6$\times$10$^6$ M$_\odot$ for Dw3 and Dw4 respectively.

\begin{figure*}
 \begin{center}
 \includegraphics[width=12cm]{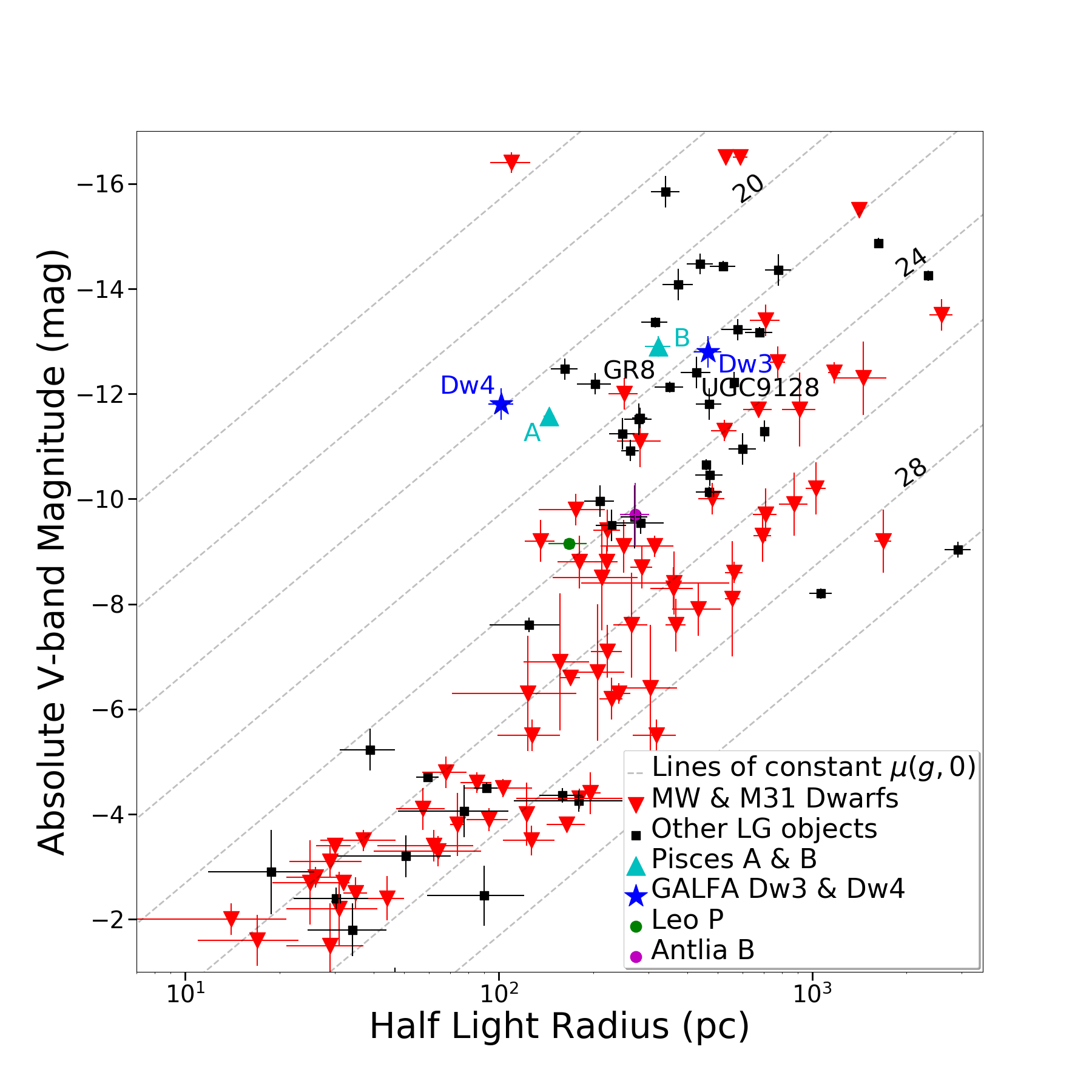}
 \caption{ Absolute V-band magnitude as a function of half-light radius for GALFA Dw3 and Dw4 (blue stars) as compared to satellites of the MW and M31 (Red Inverted Triangles) and other Local Group objects, i.e. those outside the virial radius of either the MW or M31 (Black Squares). Pisces A \& B are shown for comparison (Cyan Triangles), along with Leo P (Green Circle). The lines of constant central surface brightness assume an exponential profile and range from 16 $\textrm{mag/arcsec}^{2}$ to 30 $\textrm{mag/arcsec}^{2}$ with a line every $\Delta$2 $\textrm{mag}/\textrm{arcsec}^{2}$. 
 \label{fig:size_lum}}
 \end{center}
\end{figure*}

Our two targets broadly fit on the Local Group size-luminosity relations with slightly higher than typical surface brightness (see Figure \ref{fig:size_lum}). These properties are very similar to those found for Pisces A \& B, two other gas-rich dwarf galaxies initially found in the GALFA survey of HI compact objects \citep[][]{tollerud15,sand15}. Dw3 fits closer with the Local Group size-luminosity relation and has similar properties to many objects within the Local Group that are not satellites of the MW or M31. 
Dw4 appears to be higher surface brightness than many of these objects and is the most compact object at its magnitude \citep[][]{McConnachie18}, but has possible analogues at the edge of the Local Group such as GR8 \citep{dohm-palmer98,Tolstoy99}. This higher surface brightness when compared to Local Group satellites is likely explained by the recent star formation in both objects. These comparisons are discussed further in \S\ref{subsec:analog}.

\floattable
\begin{deluxetable}{c|cc}
\tablecaption{Properties of GALFA Dw3 \& Dw4 \label{tab:prop}}
\tablehead{
\colhead{} & \colhead{GALFA Dw3} & \colhead{GALFA Dw4} }
\startdata
R.A. (J2000) & 02$^{h}$:58$^{m}$:56$^{s}$.5$\pm$0.6 & 05$^{h}$:45$^{m}$:44$^{s}$.7$\pm$0.5 \\
Dec (J2000) & +13$^{\circ}$:37$^{'}$:45$^{''}$.4$\pm$0.5 & +10$^{\circ}$:46$^{'}$:15$^{''}$.7$\pm$0.3 \\
l (deg) & 164.15 & 195.67 \\
b (deg) & $-$38.84 & $-$24.70 \\
GALFA ID & 044.7+13.6+528 & 086.4+10.8+611 \\
Distance Modulus (mag) & 29.41$\pm$0.08 & 27.46$\pm$0.12 \\
Distance (Mpc) & 7.61$_{-0.29}^{+0.28}$ & 3.10$_{-0.17}^{+0.16}$ \\
m$_V$ (mag)\tablenotemark{a} & 16.6$\pm$0.2 & 15.7$\pm$0.2 \\
M$_V$ (mag)\tablenotemark{a} & $-$12.8$\pm$0.3 & $-$11.8$\pm$0.3 \\
V$-$I (mag) & 0.44 & 0.72 \\
E(B$-$V)\tablenotemark{b} & 0.134 & 0.531 \\
A$_{F606W}$\tablenotemark{b} & 0.322 & 1.334 \\
A$_{F814W}$\tablenotemark{b} & 0.207 & 0.811 \\
r$_h$ (\arcsec) & 12.62$\pm$1.2 & 6.82$\pm$0.06 \\
r$_h$ (pc) & 466$\pm$46 & 102$\pm$9 \\
Ellipticity & 0.54$\pm$0.03 & 0.58$\pm$0.05 \\
Position Angle (deg) & 56.4$\pm$1.7 & 100.4$\pm$1.8 \\
f$_{H\alpha}$ (erg s$^{-1}$ cm$^{-2}$) & 0.514$\pm$0.051$\times$10$^{-14}$ & 5.221$\pm$0.110$\times$10$^{-14}$ \\ 
HI~$v_{LSR}$ (km s$^{-1}$)\tablenotemark{c} & 528.59$\pm$18.90 & 614.53$\pm$40.83 \\
H$\alpha$~$v_{LSR}$ (km s$^{-1}$)\tablenotemark{d} & 503$\pm$35 & 607$\pm$35 \\
S$_{tot}$(Jy km s$^{-1}$)\tablenotemark{c} & 0.51 & 0.53 \\
M$_\star$ (M$_\odot$) & 2.1$\times$10$^6$ & 2.6$\times$10$^6$ \\
M$_{HI}$ (M$_\odot$) & 6.9$\times$10$^6$ & 1.2$\times$10$^6$ \\
SFR$_{NUV}$ (M$_\odot$) & 8.7$\pm$2.5$\times$10$^{-3}$ & -- \\
SFR$_{FUV}$ (M$_\odot$) & 8.7$\pm$0.6$\times$10$^{-4}$ & -- \\
SFR$_{H\alpha}$ (M$_\odot$) & 3.77$\pm$0.47$\times$10$^{-4}$ & 1.37$\pm$0.15$\times$10$^{-3}$ \\
\enddata\
\tablenotetext{a}{VEGA Magnitude, derived from m$_{F606W}$ using the conversion from \citep[][]{Sahu14}}
\tablenotetext{b}{Based on the \cite{Schlafly11} dust maps}
\tablenotetext{c}{From the GALFA survey, see \cite{saul12}, using the erratum values}
\tablenotetext{d}{From \cite{sand15}}
\end{deluxetable}

\subsection{HI mass} \label{subsec:HI}

The HI mass for GALFA Dw3 and Dw4 can be calculated using the HI flux and the distances derived in Section \ref{sec:TRGB}. 
This is done via the standard equation for an optically thin gas \citep{Haynes84}: 

\begin{equation}
M_{HI}=2.356\times10^{5}(D_{HI})^{2}S_{HI} M_{\odot}
\end{equation}

\noindent where D$_{HI}$ is the distance in Mpc and S$_{HI}$ is the flux in Jy km s$^{-1}$. These values are reported in Table \ref{tab:prop}. 

We use the HI fluxes from \citep[][]{saul12}\footnote{We use the revised flux values from the erratum.} and the distances we derive here, along with the standard equation to derive HI masses for GALFA Dw3 and Dw4. 
We note that these fluxes are likely underestimated due to spatial and spectral smoothing procedures employed by \cite{saul12}. An example of this underestimation is present in the discrepant fluxes for Pisces A and B, $\sim$1.2 and $\sim$1.6 Jy km s$^{-1}$, respectively, found in \cite{tollerud15} compared to 0.445 and 0.957 Jy km s$^{-1}$ from \cite{saul12}. Nevertheless, for the purpose of this work, we carry on using the values from \cite{saul12} for Dw3 and Dw4.

Given their optical luminosities, both GALFA dwarfs are relatively gas rich, with gas mass to light ratios of $\sim$0.6 M$_\odot$/L$_\odot$ for GALFA Dw3 and $\sim$0.3 M$_\odot$/L$_\odot$ for GALFA Dw4. These values are comparable to that of star forming objects within the Local Group with similar absolute magnitudes to those of the GALFA dwarfs \citep[][]{mcconnachie12}. When we compare GALFA Dw3 and Dw4 to Pisces A and B, we find that the former have smaller gas mass to light ratios (Pisces A: $\sim$2.5 M$_\odot$/L$_\odot$, Pisces B: $\sim$2.7 M$_\odot$/L$_\odot$, \citealt{tollerud16, Tollerud18,Beale20}), though this may be due to the underestimation of the HI fluxes discussed above. 
These gas masses are similar to other isolated field objects which are gas rich and star forming \citep{Geha12,Bradford15, McQuinn20}. 

\section{Star Formation Histories} \label{sec:star_form}

It is immediately apparent from the {\it HST} images and the derived CMDs that GALFA Dw3 \& Dw4 are nearby star-forming dwarf galaxies. They have well-resolved stellar populations, both show RGBs, asymptotic giant branch (AGB) stars, red helium burning stars, blue helium burning stars, MSs, an overall irregular morphology, and HII regions with young star clusters.

We attempted to use the CMD-fitting code MATCH \citep{Dolphin02} to determine the SFHs of GALFA Dw3 and Dw4 similar to other works in the Local Volume \citep[e.g.][]{McQuinn10,Weisz11,Weisz14}. 
However, the distance to these dwarfs and the shallow nature of the CMDs meant that the results did not provide meaningful constraints on the SFH of either dwarf, other than an indication of active star formation within the past 100 Myrs. Therefore we have qualitatively analyzed each dwarf's possible SFH via comparison to the Padova isochrones \citep{Bressan12} and multi-wavelength observations, similar to other works with Local Volume low-mass dwarfs where more in depth analysis has not been possible \citep[e.g.][]{mcquinn15,McQuinn20}.

\subsection{GALFA Dw3}\label{subsec:Dw3}

\subsubsection{Isochrone Comparisons}

The CMD of GALFA Dw3 reveals a complex SFH, with both young and old stellar populations. We point the reader to the left panel of Figure~\ref{fig:gal_CMD} to guide this discussion, where we denote stars in the main body of GALFA Dw3, along with those associated with its HII region (see discussion below), and plot relevant isochrones of varying age and metallicity. 

There are several faint, blue stars (with 23 $\lesssim$F814W$_0$$\lesssim$25 and (F606W$_{0}$$-$F814W$_0$) $<$ $-$0.1 mag) that are likely young MS stars, with an approximate age of $\sim$10 Myrs. Other young MS stars are apparent at fainter magnitudes. A sequence of stars spanning the same F814W$_0$ range at slightly redder colors ((F606W$_{0}$$-$F814W$_0$) $\approx$ 0.0 mag) is likely a combination of slightly older MS stars and a blue helium burning sequence, to go hand in hand with the red helium burning sequence visible at 22 $\lesssim$F814W$_0$$\lesssim$24.5 and 0.7 $\lesssim$(F606W$_{0}$$-$F814W$_0)$ $\lesssim$ 1.0 mag. A RGB is apparent at faint magnitudes (see the TRGB at F814W$_0$=25.4 mag), likely corresponding to an ancient and metal poor stellar population ($>$10--12 Gyr, [Fe/H]$\approx$$-$1.6). Stars immediately above the TRGB may be intermediate age AGB stars, or luminous helium burning stars.

The separation of the helium burning branches is a strong indicator of metallicity, with a wider separation for more metal rich systems \citep{Radburn11}, while the length and width of the branches is a good indicator of the age of the stars \citep{McQuinn11}. Approximate properties of stellar populations can even be derived for systems with very few member stars \citep[e.g.][]{Sand17}. Using the approximate length of the red helium burning branch as a guide, we estimate a stellar population with ages between 25--100 Myrs. However, for stars older than this the red helium burning branch stars becomes hard to distinguish from AGB and RGB stars \citep{McQuinn15b}.
The blue helium burning branch shows stars with ages between 25--250 Myrs. The upper limit on the duration of this star formation is determined by the completeness of the {\it HST} data. Star formation may have happened before this estimated age, however deeper data would be required to determine this. 

The size and separation of the helium burning branches in Dw3 indicate a population with [Fe/H]$\approx -1.0$, based on an approximate match to isochrones. 
A metallicity of [Fe/H]$=-1.0$ is consistent with other galaxies of similar luminosity as Dw3 (M$_V$=$-$12.8) based on the standard luminosity--metallicity relation for Local Volume galaxies \citep{Berg12}. It is also consistent within 1$\sigma$ with the possible luminosity--metallicity relation for void galaxies \citep{Pustilnik16,Kniazev18,McQuinn20}. 

Generally, dwarf irregulars form stars in bursts \citep{Weisz11}, and this is also backed up by simulations \citep{Wheeler19}. Deeper observations would be required to distinguish 
between continuous star formation and more episodic, bursty star formation in Dw3.  
Finally, isochrone fitting in the main body (excluding the HII region) shows a well populated young MS of stars below m$_{F814W}\approx 25.5$. If this is the MS turnoff for the majority of the dwarf, it would show that star formation across most of the dwarf ceased $\sim$20 Myrs ago. 

\subsubsection{H$\alpha$ Imaging}

The H$\alpha$ imaging of GALFA Dw3 (see Section \ref{sec:data}) reveals a single HII region located at the northeast edge of the dwarf, this image is shown alongside the {\it HST} image in Figure \ref{fig:gal_Ha}. 
This H$\alpha$ imaging shows a flux of 0.514$\pm$0.051$\times$10$^{-14}$ erg s$^{-1}$ cm$^{-2}$, which combined with the distance, foreground extinction and the conversion factor from \cite{Kennicutt98} implies a star formation rate of 3.77$\pm$0.47$\times$10$^{-4}$ M$_{\odot}$ yr$^{-1}$. 

If we limit the CMD to only those stars with a spatial position consistent with this HII region, we can see that the H$\alpha$ emission may be caused by a single MS O-star with a maximum age of 5 Myrs (see Figure \ref{fig:gal_CMD} and \ref{fig:gal_Ha}).  
In this region we also see a population of lower-mass young MS stars as well as red and blue helium burning stars at higher density than across the main body of the dwarf. The RGB is at a similar density in the HII region when compared to the rest of the dwarf at a similar radius, indicating the overdensity of younger stars is not simply a result of higher overall stellar density in this region. 

We also find a point source (F814W$_0$=23.2 and F606W$_0$-F814W$_0$=$-$0.25 ) consistent with an O-star, with an O5 spectral class \citep[M$_V$=$-$5.03; see the smoothed magnitudes in Table 1 of][]{Wegner00}, outside of the HII region. As this star should be massive and young enough to drive H$\alpha$ emission, but we see no H$\alpha$ emission from its position, we can draw some conclusions. The first idea would be that this is a blended multiple star system (see the Leo P analysis in \citealt{mcquinn15}).   
If we assume equally massed component stars, then these components would be O8 (M$_V$=$-$4.3) class stars, which would still be large enough to drive H$\alpha$ emission (even an equally massed triple star system would have components large enough to produce H$\alpha$). This source may be an evolved helium burning star that due to noise has been scattered into the region of the CMD equivalent to the MS.

\subsubsection{GALEX}\label{subsubsec:GALEX}

As an additional method to determine the level and spatial position of recent star formation in GALFA Dw3, we checked the GALEX archive for the dwarf's ultraviolet emission. Dw3's position was observed by GALEX as part of the All-sky Imaging Survey (AIS, exposure time $\sim$270s). These GALEX images can be seen alongside the {\it HST} images in Figure \ref{fig:gal_3_UV}. 

The GALEX data shows diffuse NUV and FUV emission across the body of Dw3, though slightly more concentrated toward the north.  
We see some concentration of FUV emission in the HII region found in the H$\alpha$ imaging, however the majority is spread across the dwarf. This significant NUV and FUV emission confirms the conclusion from the isochrone fitting that significant star formation has occurred across the dwarf within the last 100 Myrs \citep{Calzetti13}.

The detected level of NUV emission indicates that GALFA Dw3 has had recent star formation at a rate of 8.7$\pm$2.5$\times$10$^{-3}$M$_{\odot}$ yr$^{-1}$, whereas the FUV emission indicates an order of magnitude lower star formation rate of 8.7$\pm$0.6$\times$10$^{-4}$M$_{\odot}$ yr$^{-1}$. Both star formation rates were calculated using the relevant relations from \cite{iglesias06}. These relations have been shown to be potentially unreliable in low metallicity galaxies, like GALFA Dw3 \citep{mcquinn15}; in which case the star forming rate maybe up to $\sim$1.5 times higher than indicated, although this does not effect our overall results. 
The difference between the star formation rates drawn from the NUV and FUV emission may indicate that star formation in Dw3 has decreased significantly in the last $\sim$100 Myr.  
This is reinforced by the SFR derived from the H$\alpha$ imaging above  (3.77$\pm$0.47$\times$10$^{-4}$ M$_{\odot}$ yr$^{-1}$) which is comparable to the rate derived from the FUV emission but slightly lower. 
This difference in star formation rates between the tracers examined here can be explained by their differing sensitivity to different ages of star formation.  
As NUV is equally sensitive to all star formation across the last 100 Myrs, while FUV is most sensitive to stars formed in the last 10 Myrs \citep[though there is some FUV sensitivity to populations up to 100 Myrs old,][] {Calzetti13}, and H$\alpha$ is sensitive to only star formation within the last 10 Myrs. 

The UV emission coming from across the dwarf, along with the difference between the H$\alpha$, NUV and FUV, supports the conclusion drawn from the isochrone matching: that star formation was higher and more widespread in Dw3 in the recent past ($\lesssim$100 Myr), but has now quenched across most of the dwarf, and that there is ongoing star formation only in the single HII region (in the last $\sim$10 Myr).

\subsubsection{Spatial structure}

Another diagnostic that we can use to analyze GALFA Dw3 is spatial maps, see Figure \ref{fig:star_map}. When the stars are plotted on spatial maps we can see that the MS stars are concentrated in the central regions of the dwarf, have a more elliptical distribution and are preferentially found toward the northern end of the galaxy. 
This is true for all MS stars, aside from the very brightest which are only found in the HII region. 
This is in contrast to the RGB stars which are more evenly distributed throughout the galaxy. 
The helium burning stars are also more concentrated towards the center of the dwarf when compared to the RGB stars, however the concentration is less pronounced than it is for the MS stars. 

When we examine the star positions and compare them to the multi-wavelength observations, we find a strong match between the MS stars and the NUV emission. 

\subsubsection{Summary}

GALFA Dw3 shows an underlying old ($>$10--12 Gyr) metal poor ([Fe/H]$\approx$$-$1.6) stellar population across the body of the dwarf. There are also younger stellar populations. In the CMD we find well populated red and blue helium burning branches (20--100 Myr) across the body of the dwarf, this population can also be seen in the UV emission from Dw3 (see Figure \ref{fig:gal_3_UV}). Finally we also find evidence in the CMD and H$\alpha$ emission for a very young population ($<$20 Myr) that is spatially limited to a single HII region in the northeast of the dwarf (see Figure \ref{fig:gal_Ha}). 

The differences in the spatial position and extent of the tracers of different ages of star formation can be used to reconstruct a qualitative SFH for GALFA Dw3: the star formation was at a higher level and distributed more evenly throughout the dwarf in the recent past, but is now restricted to a single HII region. 
This could indicate that GALFA Dw3 is concluding an episode of recent star formation that has now been quenched outside of the HII region. 
This interpretation appears to support the model that star formation in isolated dwarf galaxies is driven by a series of `bursts' of intense star formation, interspersed with periods of quiescence \citep{Weisz11}. In this model, galaxies go through intense bursts of active star formation which expels the HI gas through stellar feedback. This expulsion of the neutral gas causes the star formation to wane and the feedback to decrease. Without feedback, more HI gas falls onto the dwarf, producing a new episode of star formation \citep{Wheeler19}. In this case, GALFA Dw3 would be in the concluding part of such a star forming episode with the last parts of star formation from an active burst. 
More detailed HI observations may be needed to determine the position and kinematic properties of the gas, as the existing HI information from the GALFA survey is low resolution \citep{saul12}.

\subsection{GALFA Dw4}\label{subsec:Dw4}

The position of GALFA Dw4 near the galactic plane complicates creating a comprehensive SFH due to the high extinction (particularly in the UV). 

\subsubsection{Isochrone Comparisons}

The CMD of GALFA Dw4 also reveals a complex SFH, with both young and old stellar populations, however there are substantial differences between Dw3 and Dw4. We point the reader to the right panel of Figure~\ref{fig:gal_CMD} to guide this discussion, where we denote stars in the main body of GALFA Dw4, along with those associated with both of its HII regions (see discussion below), and plot relevant isochrones of varying age and metallicity. 

Isochrone matching of the red and blue helium burning branches in GALFA Dw4 indicates a metallicity of [Fe/H]$\approx -$1.0 and ages of 50-500 Myrs, based on the branches' length and separation. Similar to Dw3, the red helium burning branch shows stars with ages between 50-100 Myrs, with the blue helium burning branch showing stars from 100-500 Myrs. 
The upper age boundary is limited by the completeness of the CMD so star formation may have started even earlier than 500 Myrs ago, but this can not be determined without a deeper CMD. This metallicity is consistent within 1$\sigma$ with the luminosity--metallicity relationship for Local Volume dwarfs \citep{Berg12}. 

Isochrone matching also shows Dw4 has an ancient ($>$10--12 Gyrs), low-metallicity ([Fe/H]$\approx -$1.6) RGB.  We see some evidence for a limited metallicity spread in the RGB, with some stars being consistent with [Fe/H]$\approx -$1.0, or even slightly more metal-rich, and with most likely member stars being part of this ancient population. 

Isochrone matching of Dw4 indicates that there are relatively few young MS stars when compared to the stars of the helium burning branches. This could mean that the current star formation rate is at a lower level when compared to a few hundred Myrs ago when the stars that now make up the helium burning branches were formed. This could also be a function of the very low mass of Dw4, where even if there is active star formation very few high mass MS stars are formed and therefore the higher density of helium burning branch stars is caused by the initial stellar mass function, rather than differences in star formation rate over time. This would be similar to other isolated low mass dwarfs such as Leo P or Leoncino \citep{mcquinn15,McQuinn20}. It is also possible that the young MS stars are being missed for some reason, and this possibility will be explored below.

\subsubsection{H$\alpha$ Imaging}

The H$\alpha$ imaging (see \S\ref{sec:data}) shows that Dw4 has two HII regions, one at each end of the galaxy, which match the blue regions seen in the {\it HST} imaging (see Figure \ref{fig:gal_Ha}). We find an H$\alpha$ flux of 4.184$\pm$0.097$\times$10$^{-14}$ erg s$^{-1}$ cm$^{-2}$ for the southeast region and 1.037$\pm$0.052$\times$10$^{-14}$ erg s$^{-1}$ cm$^{-2}$ for the northwest region, for a total H$\alpha$ flux of 5.221$\pm$0.110$\times$10$^{-14}$ erg s$^{-1}$ cm$^{-2}$. Combined with the distance, foreground extinction and the conversion factor from \cite{Kennicutt98} this flux implies a star formation rate of 1.37$\pm$0.15$\times$10$^{-3}$ M$_{\odot}$ yr$^{-1}$. 

When we examine these HII regions in the CMD (see the right panel of Figure \ref{fig:gal_CMD}), we find that there are no obvious O-stars to drive the H$\alpha$ emission. This could be caused by internal extinction within Dw4, which could cause the MS O-stars to appear as stars at the upper end of the blue helium burning branch. For this to be the case, the HII regions would have to be obscured by enough dust to cause extinction of A$_{F606W}$$\approx$1.1 and A$_{F814W}$$\approx$0.7. This level of extinction is substantially higher than the internal extinction reported for other dwarf galaxies \citep{McQuinn10} and is far larger than variations in the foreground extinction. 

Another possibility is that the stars which are driving H$\alpha$ emission are visible, but are not recovered in our point source photometry because they were culled at some stage in our reductions. 
To test this possibility, a CMD for Dw4 was constructed using the DOLPHOT catalog, but with the photometric quality cuts severely relaxed.  This did not detect any sources with color and brightness consistent with MS O-stars across Dw4. We have also used the ASTs to confirm that artificial stars with properties similar to MS O-stars are successfully recovered by DOLPHOT in the HII regions of Dw4. 
We also tried a similar reduction in photometric quality cuts with the crowded photometry discussed in \S\ref{sec:data}, and this yielded a few point sources consistent with MS O-stars of the spectral classes O7-O9. These could be the source of the H$\alpha$ emission, however these poorly recovered sources are generally too blue to be MS O-stars. We have considered that these objects may be O-stars with line contamination from the HII region sufficient to move it off the MS in the CMD, however this contamination would have to be larger than expected to have the observed effect. 
On the other hand, equivalent point sources are not found in the parallel field, indicating they are unique to the dwarf. `
Therefore, it is possible these are the MS O-stars, but they are in areas of the dwarf that preclude clean photometric recovery with the present data.  

It is also possible that a combination of the above scenarios are the reason we see no MS O-stars in Dw4 despite the presence of H$\alpha$ emission. In this case, internal extinction obscures and blurs the O-stars such that they are not recovered clearly by DOLPHOT.  

The two HII regions also contain most of the lower-mass MS stars seen in Dw4 (see Figure \ref{fig:star_map}). This indicates that star formation is currently limited to these two regions. We also see overdensities of red and blue helium burning stars in the HII regions compared to the dwarf as a whole. RGB stars appear to be at a similar density in the HII regions when compared to other parts of the dwarf with similar radius, indicating the overdensities of young stars  are genuine and not caused by general stellar overdensities in these regions. 

\subsubsection{SWIFT UVOT}

As stated in \S\ref{sec:data}, GALFA Dw4 is outside of the GALEX footprint due to its proximity to the galactic plane. Therefore, to get UV information on this object, {\it Swift} UVOT \citep{swift_uvot} observations were required. These were taken as part of proposal 1417202 (P.I. L. Hagen) to observe the UV dust extinction properties in GALFA Dw4 (along with 4 other Local Volume dwarfs). 

Despite these {\it Swift} images with a reasonable total exposure time ($\sim$1100s), they show no detectable UV emission from Dw4 in any of the 3 filters examined (UVW1, UVM2, UVW2). This is likely due to the high levels of extinction around Dw4 (see Table \ref{tab:prop}). The H$\alpha$ emission from Dw4, combined with the presence of bright MS stars in the {\it HST} imaging, means it is likely that there is UV emission from Dw4, but that it is not observable with the present data due to the previously mentioned high levels of extinction.

\subsubsection{Spatial structure}

In Dw4 the RGB stars are spread throughout the dwarf while the young MS and helium burning stars are largely confined to regions near the HII regions.  These younger stellar populations are at higher relative density at either end of the dwarf near the HII regions, see Figure \ref{fig:star_map}. 
We find that older helium burning stars are more evenly spread throughout the dwarf, though still more concentrated towards the current HII regions than the RGB stars. This may be the result of previous star formation being more evenly distributed, or a result of these older stars having had time to mix through the dwarf since they formed.  

\subsubsection{Summary}

GALFA Dw4 has an old ($\gtrsim$10--12 Gyrs) metal poor ([Fe/H]$\approx -$1.6) stellar population, with some evidence for a metallicity spread in the RGB. We also see younger stellar populations, with well populated red and blue helium burning sequences, and young MS stars. This is supported by H$\alpha$ imaging which shows emission concentrated in two regions at either end of the dwarf  at the same position as the young stellar populations in the CMD. 
Therefore we conclude that star formation in Dw4 is limited to the HII regions at either end of the dwarf. We also find that star formation has been ongoing for $>$500 Myrs, and seems to be more concentrated in the HII regions. 
This can be seen by the concentration of young stars in these regions compared to the RGB stars, along with the H$\alpha$ emission. However, our conclusions here are less robust than for Dw3. This is due to the lack of UV information, 
and the lower total number of stars in Dw4  
which makes it difficult to derive concrete information via examining stellar populations.

\section{Discussion} \label{sec:disc}

Having determined the distance (\S \ref{sec:TRGB}), structural properties (\S \ref{sec:structure}) and qualitative SFHs (\S \ref{sec:star_form}) of both GALFA Dw3 and Dw4, we are in a position to discuss these galaxies in detail.

\subsection{Environment} \label{subsec:environment}

We began exploring the environment around both GALFA Dw3 \& Dw4 using their newly derived distances and a search of the NASA Extragalactic Database  (NED)\footnote{http://ned.ipac.caltech.edu/}.  We searched for any catalogued objects within $\sim$5 degrees of angular separation and a relative velocity difference between $-$400 to +600  km s$^{-1}$ (this range was chosen to avoid contamination by Galactic objects with velocities less than the MW escape velocity).  This search showed that both GALFA Dw3 \& Dw4 are extremely isolated, confirming the result from \cite{sand15}.
In addition, catalogs of known galaxies were searched for objects nearby to either galaxy and we found nothing within 1.5 Mpc of either dwarf \citep[][]{Karachentsev13}. 

The closest known object to GALFA Dw3 is NGC1156: NGC1156 has a distance consistent with GALFA Dw3 at 7.6$\pm$0.7 Mpc \citep[][]{Kim12}, however with a projected separation of 11.61 degrees (1.54 Mpc at the distance of Dw3/NGC1156) and a velocity separation of 155 km s$^{-1}$ \citep{Karachentsev13}, we consider direct association at the present time to be unlikely.

GALFA Dw4 is projected near to the Orion Dwarf and A0554, however these objects are more distant at D$\sim$6.8 Mpc \citep{Anand19} and D$\sim$5.5 Mpc \citep{Karachentsev96} respectively, and therefore we consider association to be unlikely. The closest object to GALFA Dw4 is the HI source HIPASS J0630+08, with an angular separation of 11.2 degrees (a projected separation of 0.78 Mpc at the distance of Dw4)  and a velocity difference of 240 km s$^{-1}$ \citep{Karachentsev13}. This is a HI source with no detected optical counterpart \citep{Donley05}. We find that A0554 is the closest object with an optical counterpart, though this is extremely distant with a radial separation of 2.3 Mpc and a projected separation of 220 kpc. However, as GALFA Dw4 is in the `zone of avoidance' around the Galactic plane, there have been relatively few deep wide field optical surveys done in the area, and therefore it can not be ruled out that there maybe other undetected galaxies closer than A0554.  
GALFA Dw4 is also unusual as it has large peculiar velocity $\sim$+350 km s$^{-1}$, which is unexpected for isolated systems, which tend to move with the Hubble flow \citep[][]{Anand19}.  

\begin{figure*}
 \begin{center}
 \includegraphics[width=11cm]{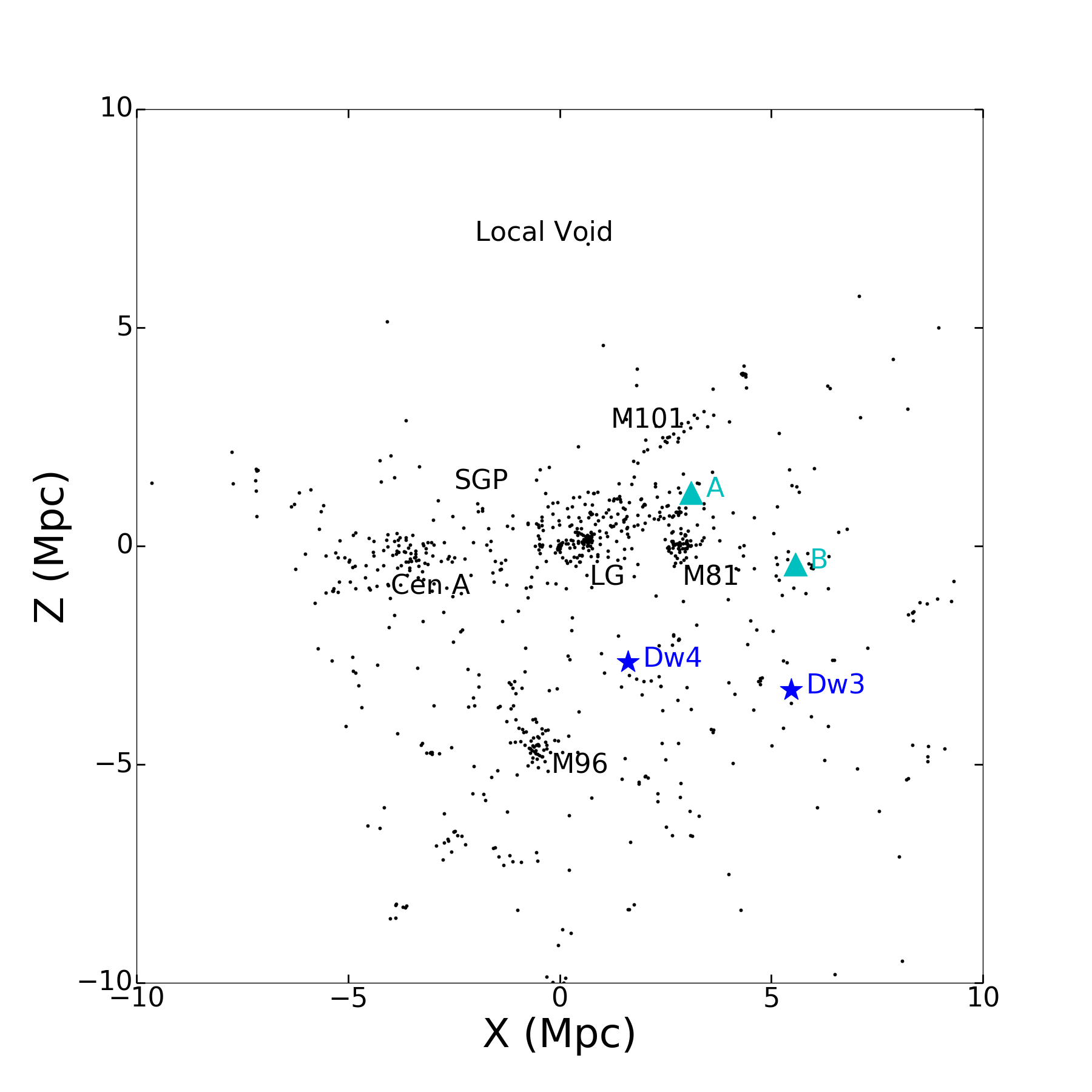}
 \includegraphics[width=11cm]{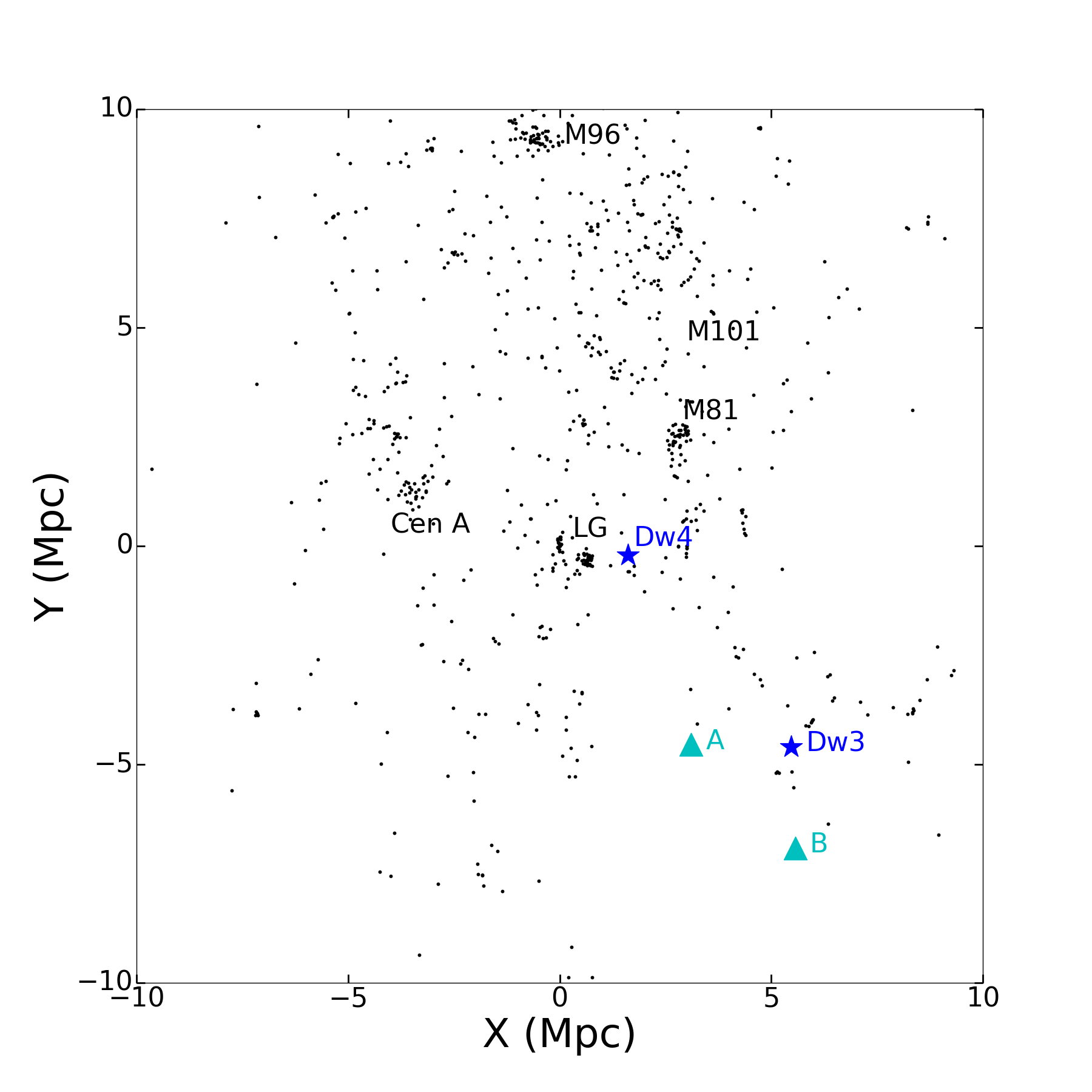}
 \caption{The location of GALFA Dw3 and Dw4 in the Local Volume. GALFA Dw3 and Dw4 are shown as blue stars and labelled, as are Pisces A and B (Cyan Triangles), while the black dots are a 10 Mpc volume-limited sample of nearby galaxies \citep{Karachentsev13}. The coordinates are supergalactic Cartesian with Earth at the center, oriented such that the x-axis points towards the origin and the z-axis points towards the Local Void \citep{Lahav00}.
 \label{fig:environment}}
 \end{center}
\end{figure*}

The isolation of GALFA Dw3 \& Dw4 can be seen in Figure \ref{fig:environment}, where the dwarfs are shown to be `below' the supergalactic plane in very low density regions of the Local Volume. 
Therefore we conclude that both GALFA Dw3 \& Dw4 are truly isolated with no other objects close enough to influence them at the current time or in the recent past. This isolation allows us to use them as probes into how star formation and galaxy evolution occur in isolated low-mass galaxies. 

\subsection{Local Volume Analogs}\label{subsec:analog}

We have examined other Local Volume dwarf galaxies to compare the properties of GALFA Dw3 and Dw4 with other low mass systems. 

GALFA Dw3 \& Dw4 have very similar physical properties to Pisces A \& B, which were also found in follow-up to the GALFA survey \citep{tollerud15,sand15}. 
All of these objects are very isolated, however Pisces A and B were theorised to be falling into local filamentary structure after spending most of cosmic time at the edge of the Local Void \citep{tollerud16}, which is speculated to have triggered recent star formation in Pisces A \& B. 

The other object from \cite{sand15}, ALFALFA Dw1 \citep[also referred to as AGC 226067 or SECCO 1;][]{bellazzini15} shows stellar populations that were found to be approximately consistent with a single burst of star formation with an age range of $\sim$7--50 Myr \citep{Sand17}, with no accompanying old stellar population, the latter being typical in almost all known dwarf galaxies. 
Based on this and other results in the literature on this object \citep{bellazzini15, Adams15,Beccari16}, there is circumstantial evidence that ALFALFA Dw1 is a distant star-forming remnant of a ram pressure stripping event in the M86 subgroup, as recent simulations have predicted \citep{Kapferer09,Tonnesen12}, and is therefore a very different class of object; it is possible that similar systems will be minor contaminants in field dwarf searches.

In Figure \ref{fig:size_lum}, there are a number of Local Volume objects that have similar physical properties to GALFA Dw3 and Dw4. 
UGC9128 is an isolated Local Volume object \citep[D$\sim$2.3 Mpc;][]{tully13}, and is a good analog for Dw3. It has very similar physical properties and a recent SFH that is comparable to Dw3, with recent star formation throughout the dwarf but current star formation limited to a few small regions \citep{McQuinn10}. UGC9128 shows evidence of having had 3 bursts of star formation in the last $\sim$500 Myrs \citep{McQuinn10}.  
GR8 (DDO155/UGC8091) is a star forming dwarf in the Local Volume  
with a distance of $\sim$2.2 Mpc \citep{tully13}. It has very similar physical properties to GALFA Dw4.  
In GR8, star formation is limited to HII complexes which seem to arise in associated regions approximately 100-200 pc in size, which last for $\sim$100 Myrs before star formation ceases and new regions begin to actively form stars \citep{dohm-palmer98, Tolstoy99}. 

The Survey of HI in Extremely Low-mass Dwarfs (SHIELD) galaxies \citep{Cannon11} are a selection of 12 galaxies initially detected in ALFALFA \citep{Giovanelli05,Haynes18} data. These galaxies were selected based on low HI and stellar mass estimates. In terms of absolute magnitude and gas mass, the SHIELD galaxies are in the same range as GALFA Dw3 and Dw4.  
Examination of the SFHs of the SHIELD galaxies also shows a recent star formation rate consistent with that derived for Dw3 (see \S\ref{subsubsec:GALEX}). 
The SHIELD galaxies are found in a number of different environments, with three (AGC 748778, AGC 174605, and AGC 74923) being isolated \citep[$>$1 Mpc from their nearest neighbors,][]{McQuinn14,McQuinn15b}. These objects have very similar physical properties to Dw3, making them potentially good analogs, while Dw4 is fainter and physically smaller than the typical SHIELD galaxy. As previously mentioned Dw4 is one of the most compact objects at its luminosity detected.

\section{Conclusions} \label{sec:conclusion}

We have presented {\it HST} imaging of GALFA Dw3 and Dw4, two Local Volume dwarf galaxies which were initially discovered as optical counterparts to compact HI clouds in the GALFA survey.  Both dwarfs resolve into stars, displaying complex stellar populations, including an old red giant branch, young helium burning sequences and main sequence stars.  Each system also has young star clusters and HII regions which are evident in our H$\alpha$ imaging.  In detail, the two dwarfs appear to have slightly different star formation histories based on a qualitative assessment of their CMDs and on the available UV data.  GALFA Dw3 shows signs of a recently ceased episode of active star formation; although it is not well constrained, Dw4 seems to have a more consistent level of star formation within spatially limited HII regions at either end of the dwarf. 

Using the resolved CMDs, we measure the distance to each dwarf using the TRGB method, finding $D$=7.61$_{-0.29}^{+0.28}$ Mpc and $D$=3.10$_{-0.17}^{+0.16}$ Mpc for GALFA Dw3 and Dw4, respectively.  With this information in hand, we found each dwarf to be extremely isolated, with no known neighbor within $\sim$1.5 Mpc, suggesting that neither galaxy has experienced a significant environmental influence.

GALFA Dw3 and Dw4 are similar to other Local Volume dwarfs initially detected in wide field HI surveys \citep[see \S\ref{subsec:analog}][]{Cannon11,tollerud15,sand15}. 
The lack of detections of new gas-rich low-mass dwarf galaxies within the Local Group \citep[similar to Leo P or Leo T][]{Irwin07,rhode13} in these surveys indicates that these `mini-halos' are likely rare. 
The lack of new Local Group objects found in the GALFA survey has been used to examine a potential link between HI gas in dwarfs and the lower mass limit for reionization \citep{Tollerud18}. 
It found that the lack of detections was very unlikely if these objects were common and this rarity could be used to determine the lower mass limit for reionization \citep[see][for more details]{Tollerud18}. 

GALFA Dw3 and Dw4 (and other systems like them, such as Pisces A and B; \citealt{tollerud16}) present a unique opportunity to examine low-metallicity isolated dwarf galaxies analogous  to the earliest galaxies in the Universe.  Further work on GALFA Dw3 and Dw4, and related objects, will include gas phase metallicity measurements \citep[e.g.][]{Hirschauer16,McQuinn20} and high resolution HI mapping \citep[e.g.][]{Beale20} to further understand the driving mechanisms of the structure and evolution of the faintest dwarf galaxies.

\acknowledgments

Research by PB is supported by NASA through grant number HST-GO-14796.005-A from the Space Telescope Science Institute which is operated by AURA, Inc., under NASA contract NAS 5-26555. Research by DJS is supported by NSF grants  AST-1821967 and AST-1813708. Research by DC is supported by NSF grant AST-1814208, and by NASA through grants number HST-GO-15426.007-A and HST-GO-15332.004-A from the Space Telescope Science Institute, which is operated by AURA, Inc., under NASA contract NAS 5-26555. BMP is supported by an NSF Astronomy and Astrophysics Postdoctoral Fellowship under award AST-2001663. EO is partially supported by NSF grant AST-1815767. JS acknowledges support from the Packard Foundation.

This publication utilizes data from Galactic ALFA HI (GALFA HI) survey data set obtained with the Arecibo L-band Feed Array (ALFA) on the Arecibo 305m telescope. The Arecibo Observatory is operated by SRI International under a cooperative agreement with the National Science Foundation (AST-1100968), and in alliance with Ana G. M\'{e}ndez-Universidad Metropolitana, and the Universities Space Research Association. The GALFA HI surveys have been funded by the NSF through grants to Columbia University, the University of Wisconsin, and the University of California. 

\facilities{HST (ACS), WIYN:0.9m, GALEX, SWIFT} 
\software{Numpy, Astropy \citep{astropy}, DOLPHOT \citep{dolphin00}}

\bibliographystyle{aasjournal}
\bibliography{ref_PB}

\begin{thebibliography}{}
\expandafter\ifx\csname natexlab\endcsname\relax\def\natexlab#1{#1}\fi
\providecommand{\url}[1]{\href{#1}{#1}}

\bibitem[{{Adams} {et~al.}(2013){Adams}, {Giovanelli}, \& {Haynes}}]{adams13}
{Adams}, E.~A.~K., {Giovanelli}, R., \& {Haynes}, M.~P. 2013, \apj, 768, 77

\bibitem[{{Adams} {et~al.}(2015){Adams}, {Faerman}, {Janesh}, {Janowiecki},
  {Oosterloo}, {Rhode}, {Giovanelli}, {Haynes}, {Salzer}, {Sternberg},
  {Cannon}, \& {Mu{\~n}oz}}]{Adams15}
{Adams}, E.~A.~K., {Faerman}, Y., {Janesh}, W.~F., {et~al.} 2015, \aap, 573, L3

\bibitem[{{Anand} {et~al.}(2019){Anand}, {Tully}, {Rizzi}, {Shaya}, \&
  {Karachentsev}}]{Anand19}
{Anand}, G.~S., {Tully}, R.~B., {Rizzi}, L., {Shaya}, E.~J., \& {Karachentsev},
  I.~D. 2019, \apj, 880, 52

\bibitem[{{Beale} {et~al.}(2020){Beale}, {Donovan Meyer}, {Tollerud}, {Putman},
  \& {Peek}}]{Beale20}
{Beale}, L., {Donovan Meyer}, J., {Tollerud}, E.~J., {Putman}, M.~E., \&
  {Peek}, J.~E.~G. 2020, arXiv e-prints, arXiv:2009.09145

\bibitem[{{Beccari} {et~al.}(2016){Beccari}, {Bellazzini}, {Battaglia},
  {Ibata}, {Martin}, {Testa}, {Cignoni}, \& {Correnti}}]{Beccari16}
{Beccari}, G., {Bellazzini}, M., {Battaglia}, G., {et~al.} 2016, \aap, 591, A56

\bibitem[{{Bell} \& {de Jong}(2001)}]{Bell01}
{Bell}, E.~F., \& {de Jong}, R.~S. 2001, \apj, 550, 212

\bibitem[{{Bellazzini} {et~al.}(2015){Bellazzini}, {Magrini}, {Mucciarelli},
  {Beccari}, {Ibata}, {Battaglia}, {Martin}, {Testa}, {Fumana}, {Marchetti},
  {Correnti}, \& {Fraternali}}]{bellazzini15}
{Bellazzini}, M., {Magrini}, L., {Mucciarelli}, A., {et~al.} 2015, \apjl, 800,
  L15

\bibitem[{{Bennet} {et~al.}(2019){Bennet}, {Sand}, {Crnojevi{\'c}}, {Spekkens},
  {Karunakaran}, {Zaritsky}, \& {Mutlu-Pakdil}}]{Bennet19}
{Bennet}, P., {Sand}, D.~J., {Crnojevi{\'c}}, D., {et~al.} 2019, \apj, 885, 153

\bibitem[{{Bennet} {et~al.}(2020){Bennet}, {Sand}, {Crnojevi{\'c}}, {Spekkens},
  {Karunakaran}, {Zaritsky}, \& {Mutlu-Pakdil}}]{Bennet20}
---. 2020, \apjl, 893, L9

\bibitem[{{Berg} {et~al.}(2012){Berg}, {Skillman}, {Marble}, {van Zee},
  {Engelbracht}, {Lee}, {Kennicutt}, {Calzetti}, {Dale}, \& {Johnson}}]{Berg12}
{Berg}, D.~A., {Skillman}, E.~D., {Marble}, A.~R., {et~al.} 2012, \apj, 754, 98

\bibitem[{{Bradford} {et~al.}(2015){Bradford}, {Geha}, \&
  {Blanton}}]{Bradford15}
{Bradford}, J.~D., {Geha}, M.~C., \& {Blanton}, M.~R. 2015, \apj, 809, 146

\bibitem[{{Bressan} {et~al.}(2012){Bressan}, {Marigo}, {Girardi}, {Salasnich},
  {Dal Cero}, {Rubele}, \& {Nanni}}]{Bressan12}
{Bressan}, A., {Marigo}, P., {Girardi}, L., {et~al.} 2012, \mnras, 427, 127

\bibitem[{{Bullock} \& {Boylan-Kolchin}(2017)}]{Bullock17}
{Bullock}, J.~S., \& {Boylan-Kolchin}, M. 2017, Annual Review of Astronomy and
  Astrophysics, 55, 343

\bibitem[{{Calzetti}(2013)}]{Calzetti13}
{Calzetti}, D. 2013, {Star Formation Rate Indicators}, ed.
  J.~{Falc{\'o}n-Barroso} \& J.~H. {Knapen}, 419

\bibitem[{{Cannon} {et~al.}(2011){Cannon}, {Giovanelli}, {Haynes},
  {Janowiecki}, {Parker}, {Salzer}, {Adams}, {Engstrom}, {Huang}, {McQuinn},
  {Ott}, {Saintonge}, {Skillman}, {Allan}, {Erny}, {Fliss}, \&
  {Smith}}]{Cannon11}
{Cannon}, J.~M., {Giovanelli}, R., {Haynes}, M.~P., {et~al.} 2011, \apjl, 739,
  L22

\bibitem[{{Carlsten} {et~al.}(2020){Carlsten}, {Greco}, {Beaton}, \&
  {Greene}}]{Carlsten20}
{Carlsten}, S.~G., {Greco}, J.~P., {Beaton}, R.~L., \& {Greene}, J.~E. 2020,
  \apj, 891, 144

\bibitem[{{Crnojevi{\'c}} {et~al.}(2019){Crnojevi{\'c}}, {Sand}, {Bennet},
  {Pasetto}, {Spekkens}, {Caldwell}, {Guhathakurta}, {McLeod}, {Seth}, {Simon},
  {Strader}, \& {Toloba}}]{Crnojevic19}
{Crnojevi{\'c}}, D., {Sand}, D.~J., {Bennet}, P., {et~al.} 2019, \apj, 872, 80

\bibitem[{{Da Costa} \& {Armandroff}(1990)}]{dacosta90}
{Da Costa}, G.~S., \& {Armandroff}, T.~E. 1990, \aj, 100, 162

\bibitem[{{Dohm-Palmer} {et~al.}(1998){Dohm-Palmer}, {Skillman}, {Gallagher},
  {Tolstoy}, {Mateo}, {Dufour}, {Saha}, {Hoessel}, \& {Chiosi}}]{dohm-palmer98}
{Dohm-Palmer}, R.~C., {Skillman}, E.~D., {Gallagher}, J., {et~al.} 1998, \aj,
  116, 1227

\bibitem[{{Dolphin}(2000)}]{dolphin00}
{Dolphin}, A.~E. 2000, \pasp, 112, 1383

\bibitem[{{Dolphin}(2002)}]{Dolphin02}
---. 2002, \mnras, 332, 91

\bibitem[{{Donley} {et~al.}(2005){Donley}, {Staveley-Smith}, {Kraan-Korteweg},
  {Islas-Islas}, {Schr{\"o}der}, {Henning}, {Koribalski}, {Mader}, \&
  {Stewart}}]{Donley05}
{Donley}, J.~L., {Staveley-Smith}, L., {Kraan-Korteweg}, R.~C., {et~al.} 2005,
  \aj, 129, 220

\bibitem[{{Freedman} {et~al.}(2020){Freedman}, {Madore}, {Hoyt}, {Jang},
  {Beaton}, {Lee}, {Monson}, {Neeley}, \& {Rich}}]{Freedman20}
{Freedman}, W.~L., {Madore}, B.~F., {Hoyt}, T., {et~al.} 2020, \apj, 891, 57

\bibitem[{{Garrison-Kimmel} {et~al.}(2014){Garrison-Kimmel}, {Boylan-Kolchin},
  {Bullock}, \& {Lee}}]{Garrison14}
{Garrison-Kimmel}, S., {Boylan-Kolchin}, M., {Bullock}, J.~S., \& {Lee}, K.
  2014, \mnras, 438, 2578

\bibitem[{{Garrison-Kimmel} {et~al.}(2019){Garrison-Kimmel}, {Wetzel},
  {Hopkins}, {Sanderson}, {El-Badry}, {Graus}, {Chan}, {Feldmann},
  {Boylan-Kolchin}, {Hayward}, {Bullock}, {Fitts}, {Samuel}, {Wheeler},
  {Kere{\v{s}}}, \& {Faucher-Gigu{\`e}re}}]{Garrison19}
{Garrison-Kimmel}, S., {Wetzel}, A., {Hopkins}, P.~F., {et~al.} 2019, \mnras,
  489, 4574

\bibitem[{{Geha} {et~al.}(2012){Geha}, {Blanton}, {Yan}, \& {Tinker}}]{Geha12}
{Geha}, M., {Blanton}, M.~R., {Yan}, R., \& {Tinker}, J.~L. 2012, \apj, 757, 85

\bibitem[{{Gehrels} {et~al.}(2004){Gehrels}, {Chincarini}, {Giommi}, {Mason},
  {Nousek}, {Wells}, {White}, {Barthelmy}, {Burrows}, {Cominsky}, {Hurley},
  {Marshall}, {M{\'e}sz{\'a}ros}, {Roming}, {Angelini}, {Barbier}, {Belloni},
  {Campana}, {Caraveo}, {Chester}, {Citterio}, {Cline}, {Cropper}, {Cummings},
  {Dean}, {Feigelson}, {Fenimore}, {Frail}, {Fruchter}, {Garmire}, {Gendreau},
  {Ghisellini}, {Greiner}, {Hill}, {Hunsberger}, {Krimm}, {Kulkarni}, {Kumar},
  {Lebrun}, {Lloyd-Ronning}, {Markwardt}, {Mattson}, {Mushotzky}, {Norris},
  {Osborne}, {Paczynski}, {Palmer}, {Park}, {Parsons}, {Paul}, {Rees},
  {Reynolds}, {Rhoads}, {Sasseen}, {Schaefer}, {Short}, {Smale}, {Smith},
  {Stella}, {Tagliaferri}, {Takahashi}, {Tashiro}, {Townsley}, {Tueller},
  {Turner}, {Vietri}, {Voges}, {Ward}, {Willingale}, {Zerbi}, \&
  {Zhang}}]{Gehrels04}
{Gehrels}, N., {Chincarini}, G., {Giommi}, P., {et~al.} 2004, \apj, 611, 1005

\bibitem[{{Giovanelli} {et~al.}(2010){Giovanelli}, {Haynes}, {Kent}, \&
  {Adams}}]{Giovanelli10}
{Giovanelli}, R., {Haynes}, M.~P., {Kent}, B.~R., \& {Adams}, E. A.~K. 2010,
  \apjl, 708, L22

\bibitem[{{Giovanelli} {et~al.}(2005){Giovanelli}, {Haynes}, {Kent},
  {Perillat}, {Saintonge}, {Brosch}, {Catinella}, {Hoffman}, {Stierwalt},
  {Spekkens}, {Lerner}, {Masters}, {Momjian}, {Rosenberg}, {Springob},
  {Boselli}, {Charmand aris}, {Darling}, {Davies}, {Garcia Lambas}, {Gavazzi},
  {Giovanardi}, {Hardy}, {Hunt}, {Iovino}, {Karachentsev}, {Karachentseva},
  {Koopmann}, {Marinoni}, {Minchin}, {Muller}, {Putman}, {Pantoja}, {Salzer},
  {Scodeggio}, {Skillman}, {Solanes}, {Valotto}, {van Driel}, \& {van
  Zee}}]{Giovanelli05}
{Giovanelli}, R., {Haynes}, M.~P., {Kent}, B.~R., {et~al.} 2005, \aj, 130, 2598

\bibitem[{{Haynes} \& {Giovanelli}(1984)}]{Haynes84}
{Haynes}, M.~P., \& {Giovanelli}, R. 1984, \aj, 89, 758

\bibitem[{{Haynes} {et~al.}(2018){Haynes}, {Giovanelli}, {Kent}, {Adams},
  {Balonek}, {Craig}, {Fertig}, {Finn}, {Giovanardi}, {Hallenbeck}, {Hess},
  {Hoffman}, {Huang}, {Jones}, {Koopmann}, {Kornreich}, {Leisman}, {Miller},
  {Moorman}, {O'Connor}, {O'Donoghue}, {Papastergis}, {Troischt}, {Stark}, \&
  {Xiao}}]{Haynes18}
{Haynes}, M.~P., {Giovanelli}, R., {Kent}, B.~R., {et~al.} 2018, \apj, 861, 49

\bibitem[{{Hirschauer} {et~al.}(2016){Hirschauer}, {Salzer}, {Skillman},
  {Berg}, {McQuinn}, {Cannon}, {Gordon}, {Haynes}, {Giovanelli}, {Adams},
  {Janowiecki}, {Rhode}, {Pogge}, {Croxall}, \& {Aver}}]{Hirschauer16}
{Hirschauer}, A.~S., {Salzer}, J.~J., {Skillman}, E.~D., {et~al.} 2016, \apj,
  822, 108

\bibitem[{{Iglesias-P{\'a}ramo} {et~al.}(2006){Iglesias-P{\'a}ramo}, {Buat},
  {Takeuchi}, {Xu}, {Boissier}, {Boselli}, {Burgarella}, {Madore}, {Gil de
  Paz}, {Bianchi}, {Barlow}, {Byun}, {Donas}, {Forster}, {Friedman}, {Heckman},
  {Jelinski}, {Lee}, {Malina}, {Martin}, {Milliard}, {Morrissey}, {Neff},
  {Rich}, {Schiminovich}, {Seibert}, {Siegmund}, {Small}, {Szalay}, {Welsh}, \&
  {Wyder}}]{iglesias06}
{Iglesias-P{\'a}ramo}, J., {Buat}, V., {Takeuchi}, T.~T., {et~al.} 2006, \apjs,
  164, 38

\bibitem[{{Irwin} {et~al.}(2007){Irwin}, {Belokurov}, {Evans}, {Ryan-Weber},
  {de Jong}, {Koposov}, {Zucker}, {Hodgkin}, {Gilmore}, {Prema}, {Hebb},
  {Begum}, {Fellhauer}, {Hewett}, {Kennicutt}, {Wilkinson}, {Bramich},
  {Vidrih}, {Rix}, {Beers}, {Barentine}, {Brewington}, {Harvanek},
  {Krzesinski}, {Long}, {Nitta}, \& {Snedden}}]{Irwin07}
{Irwin}, M.~J., {Belokurov}, V., {Evans}, N.~W., {et~al.} 2007, \apjl, 656, L13

\bibitem[{{Jang} \& {Lee}(2017)}]{jang17}
{Jang}, I.~S., \& {Lee}, M.~G. 2017, \apj, 835, 28

\bibitem[{{Kapferer} {et~al.}(2009){Kapferer}, {Sluka}, {Schindler}, {Ferrari},
  \& {Ziegler}}]{Kapferer09}
{Kapferer}, W., {Sluka}, C., {Schindler}, S., {Ferrari}, C., \& {Ziegler}, B.
  2009, \aap, 499, 87

\bibitem[{{Karachentsev} \& {Musella}(1996)}]{Karachentsev96}
{Karachentsev}, I., \& {Musella}, I. 1996, \aap, 315, 348

\bibitem[{{Karachentsev} {et~al.}(2013){Karachentsev}, {Makarov}, \&
  {Kaisina}}]{Karachentsev13}
{Karachentsev}, I.~D., {Makarov}, D.~I., \& {Kaisina}, E.~I. 2013, \aj, 145,
  101

\bibitem[{{Kennicutt}(1998)}]{Kennicutt98}
{Kennicutt}, Robert~C., J. 1998, \araa, 36, 189

\bibitem[{{Kim} {et~al.}(2012){Kim}, {Park}, {Kyeong}, {Lee}, {Ree}, \&
  {Kim}}]{Kim12}
{Kim}, S.~C., {Park}, H.~S., {Kyeong}, J., {et~al.} 2012, \pasj, 64, 23

\bibitem[{{Kniazev} {et~al.}(2018){Kniazev}, {Egorova}, \&
  {Pustilnik}}]{Kniazev18}
{Kniazev}, A.~Y., {Egorova}, E.~S., \& {Pustilnik}, S.~A. 2018, \mnras, 479,
  3842

\bibitem[{{Lahav} {et~al.}(2000){Lahav}, {Santiago}, {Webster}, {Strauss},
  {Davis}, {Dressler}, \& {Huchra}}]{Lahav00}
{Lahav}, O., {Santiago}, B.~X., {Webster}, A.~M., {et~al.} 2000, \mnras, 312,
  166

\bibitem[{{Lee} {et~al.}(1993){Lee}, {Freedman}, \& {Madore}}]{lee93}
{Lee}, M.~G., {Freedman}, W.~L., \& {Madore}, B.~F. 1993, \apj, 417, 553

\bibitem[{{Makarov} {et~al.}(2006){Makarov}, {Makarova}, {Rizzi}, {Tully},
  {Dolphin}, {Sakai}, \& {Shaya}}]{Makarov06}
{Makarov}, D., {Makarova}, L., {Rizzi}, L., {et~al.} 2006, \aj, 132, 2729

\bibitem[{{Mao} {et~al.}(2020){Mao}, {Geha}, {Wechsler}, {Weiner}, {Tollerud},
  {Nadler}, \& {Kallivayalil}}]{Mao20}
{Mao}, Y.-Y., {Geha}, M., {Wechsler}, R.~H., {et~al.} 2020, arXiv e-prints,
  arXiv:2008.12783

\bibitem[{{Martin} \& {GALEX Team}(2005)}]{martin05}
{Martin}, C., \& {GALEX Team}. 2005, in IAU Symposium, Vol. 216, Maps of the
  Cosmos, ed. M.~{Colless}, L.~{Staveley-Smith}, \& R.~A. {Stathakis}, 221

\bibitem[{{Martin} {et~al.}(2008){Martin}, {de Jong}, \& {Rix}}]{Martin08}
{Martin}, N.~F., {de Jong}, J. T.~A., \& {Rix}, H.-W. 2008, \apj, 684, 1075

\bibitem[{{McConnachie}(2012)}]{mcconnachie12}
{McConnachie}, A.~W. 2012, \aj, 144, 4

\bibitem[{{McConnachie} {et~al.}(2018){McConnachie}, {Ibata}, {Martin},
  {Ferguson}, {Collins}, {Gwyn}, {Irwin}, {Lewis}, {Mackey}, {Davidge},
  {Arias}, {Conn}, {C{\^o}t{\'e}}, {Crnojevic}, {Huxor}, {Penarrubia},
  {Spengler}, {Tanvir}, {Valls-Gabaud}, {Babul}, {Barmby}, {Bate}, {Bernard},
  {Chapman}, {Dotter}, {Harris}, {McMonigal}, {Navarro}, {Puzia}, {Rich},
  {Thomas}, \& {Widrow}}]{McConnachie18}
{McConnachie}, A.~W., {Ibata}, R., {Martin}, N., {et~al.} 2018, \apj, 868, 55

\bibitem[{{McQuinn} {et~al.}(2011){McQuinn}, {Skillman}, {Dalcanton},
  {Dolphin}, {Holtzman}, {Weisz}, \& {Williams}}]{McQuinn11}
{McQuinn}, K. B.~W., {Skillman}, E.~D., {Dalcanton}, J.~J., {et~al.} 2011,
  \apj, 740, 48

\bibitem[{{McQuinn} {et~al.}(2015{\natexlab{a}}){McQuinn}, {Skillman},
  {Dolphin}, \& {Mitchell}}]{mcquinn15}
{McQuinn}, K. B.~W., {Skillman}, E.~D., {Dolphin}, A.~E., \& {Mitchell}, N.~P.
  2015{\natexlab{a}}, \apj, 808, 109

\bibitem[{{McQuinn} {et~al.}(2010){McQuinn}, {Skillman}, {Cannon}, {Dalcanton},
  {Dolphin}, {Hidalgo-Rodr{\'\i}guez}, {Holtzman}, {Stark}, {Weisz}, \&
  {Williams}}]{McQuinn10}
{McQuinn}, K. B.~W., {Skillman}, E.~D., {Cannon}, J.~M., {et~al.} 2010, \apj,
  724, 49

\bibitem[{{McQuinn} {et~al.}(2013){McQuinn}, {Skillman}, {Berg}, {Cannon},
  {Salzer}, {Adams}, {Dolphin}, {Giovanelli}, {Haynes}, \& {Rhode}}]{mcquinn13}
{McQuinn}, K.~B.~W., {Skillman}, E.~D., {Berg}, D., {et~al.} 2013, \aj, 146,
  145

\bibitem[{{McQuinn} {et~al.}(2014){McQuinn}, {Cannon}, {Dolphin}, {Skillman},
  {Salzer}, {Haynes}, {Adams}, {Cave}, {Elson}, {Giovanelli}, {Ott}, \&
  {Saintonge}}]{McQuinn14}
{McQuinn}, K. B.~W., {Cannon}, J.~M., {Dolphin}, A.~E., {et~al.} 2014, \apj,
  785, 3

\bibitem[{{McQuinn} {et~al.}(2015{\natexlab{b}}){McQuinn}, {Cannon}, {Dolphin},
  {Skillman}, {Haynes}, {Simones}, {Salzer}, {Adams}, {Elson}, {Giovanelli}, \&
  {Ott}}]{McQuinn15b}
---. 2015{\natexlab{b}}, \apj, 802, 66

\bibitem[{{McQuinn} {et~al.}(2020){McQuinn}, {Berg}, {Skillman}, {Adams},
  {Cannon}, {Dolphin}, {Salzer}, {Giovanelli}, {Haynes}, {Hirschauer},
  {Janoweicki}, {Klapkowski}, \& {Rhode}}]{McQuinn20}
{McQuinn}, K. B.~W., {Berg}, D.~A., {Skillman}, E.~D., {et~al.} 2020, \apj,
  891, 181

\bibitem[{{Morrissey} {et~al.}(2007){Morrissey}, {Conrow}, {Barlow}, {Small},
  {Seibert}, {Wyder}, {Budav{\'a}ri}, {Arnouts}, {Friedman}, {Forster},
  {Martin}, {Neff}, {Schiminovich}, {Bianchi}, {Donas}, {Heckman}, {Lee},
  {Madore}, {Milliard}, {Rich}, {Szalay}, {Welsh}, \& {Yi}}]{Morrissey07}
{Morrissey}, P., {Conrow}, T., {Barlow}, T.~A., {et~al.} 2007, \apjs, 173, 682

\bibitem[{{Pustilnik} {et~al.}(2016){Pustilnik}, {Perepelitsyna}, \&
  {Kniazev}}]{Pustilnik16}
{Pustilnik}, S.~A., {Perepelitsyna}, Y.~A., \& {Kniazev}, A.~Y. 2016, \mnras,
  463, 670

\bibitem[{{Radburn-Smith} {et~al.}(2011){Radburn-Smith}, {de Jong}, {Seth},
  {Bailin}, {Bell}, {Brown}, {Bullock}, {Courteau}, {Dalcanton}, {Ferguson},
  {Goudfrooij}, {Holfeltz}, {Holwerda}, {Purcell}, {Sick}, {Streich}, {Vlajic},
  \& {Zucker}}]{Radburn11}
{Radburn-Smith}, D.~J., {de Jong}, R.~S., {Seth}, A.~C., {et~al.} 2011, \apjs,
  195, 18

\bibitem[{{Rhode} {et~al.}(2013){Rhode}, {Salzer}, {Haurberg}, {Van Sistine},
  {Young}, {Haynes}, {Giovanelli}, {Cannon}, {Skillman}, {McQuinn}, \&
  {Adams}}]{rhode13}
{Rhode}, K.~L., {Salzer}, J.~J., {Haurberg}, N.~C., {et~al.} 2013, \aj, 145,
  149

\bibitem[{{Rizzi} {et~al.}(2007){Rizzi}, {Tully}, {Makarov}, {Makarova},
  {Dolphin}, {Sakai}, \& {Shaya}}]{Rizzi07}
{Rizzi}, L., {Tully}, R.~B., {Makarov}, D., {et~al.} 2007, \apj, 661, 815

\bibitem[{{Roming} {et~al.}(2005){Roming}, {Kennedy}, {Mason}, {Nousek}, {Ahr},
  {Bingham}, {Broos}, {Carter}, {Hancock}, {Huckle}, {Hunsberger}, {Kawakami},
  {Killough}, {Koch}, {McLelland}, {Smith}, {Smith}, {Soto}, {Boyd},
  {Breeveld}, {Holland}, {Ivanushkina}, {Pryzby}, {Still}, \&
  {Stock}}]{swift_uvot}
{Roming}, P. W.~A., {Kennedy}, T.~E., {Mason}, K.~O., {et~al.} 2005, \ssr, 120,
  95

\bibitem[{{Sahu} {et~al.}(2014){Sahu}, {Deustua}, \& {Sabbi}}]{Sahu14}
{Sahu}, K., {Deustua}, S., \& {Sabbi}, E. 2014, {WFC3/UVIS Photometric
  Transformations}, Tech. rep.

\bibitem[{{Sand} {et~al.}(2009){Sand}, {Olszewski}, {Willman}, {Zaritsky},
  {Seth}, {Harris}, {Piatek}, \& {Saha}}]{Sand09}
{Sand}, D.~J., {Olszewski}, E.~W., {Willman}, B., {et~al.} 2009, \apj, 704, 898

\bibitem[{{Sand} {et~al.}(2012){Sand}, {Strader}, {Willman}, {Zaritsky},
  {McLeod}, {Caldwell}, {Seth}, \& {Olszewski}}]{Sand12}
{Sand}, D.~J., {Strader}, J., {Willman}, B., {et~al.} 2012, \apj, 756, 79

\bibitem[{{Sand} {et~al.}(2015){Sand}, {Crnojevi{\'c}}, {Bennet}, {Willman},
  {Hargis}, {Strader}, {Olszewski}, {Tollerud}, {Simon}, {Caldwell},
  {Guhathakurta}, {James}, {Koposov}, {McLeod}, {Morrell}, {Peacock},
  {Salinas}, {Seth}, {Stark}, \& {Toloba}}]{sand15}
{Sand}, D.~J., {Crnojevi{\'c}}, D., {Bennet}, P., {et~al.} 2015, \apj, 806, 95

\bibitem[{{Sand} {et~al.}(2017){Sand}, {Seth}, {Crnojevi{\'c}}, {Spekkens},
  {Strader}, {Adams}, {Caldwell}, {Guhathakurta}, {Kenney}, {Rand all},
  {Simon}, {Toloba}, \& {Willman}}]{Sand17}
{Sand}, D.~J., {Seth}, A.~C., {Crnojevi{\'c}}, D., {et~al.} 2017, \apj, 843,
  134

\bibitem[{{Saul} {et~al.}(2012){Saul}, {Peek}, {Grcevich}, {Putman}, {Douglas},
  {Korpela}, {Stanimirovi{\'c}}, {Heiles}, {Gibson}, {Lee}, {Begum}, {Brown},
  {Burkhart}, {Hamden}, {Pingel}, \& {Tonnesen}}]{saul12}
{Saul}, D.~R., {Peek}, J.~E.~G., {Grcevich}, J., {et~al.} 2012, \apj, 758, 44

\bibitem[{{Schlafly} \& {Finkbeiner}(2011)}]{Schlafly11}
{Schlafly}, E.~F., \& {Finkbeiner}, D.~P. 2011, \apj, 737, 103

\bibitem[{{Schlegel} {et~al.}(1998){Schlegel}, {Finkbeiner}, \&
  {Davis}}]{schlegel98}
{Schlegel}, D.~J., {Finkbeiner}, D.~P., \& {Davis}, M. 1998, \apj, 500, 525

\bibitem[{{Spekkens} {et~al.}(2014){Spekkens}, {Urbancic}, {Mason}, {Willman},
  \& {Aguirre}}]{spekkens14}
{Spekkens}, K., {Urbancic}, N., {Mason}, B.~S., {Willman}, B., \& {Aguirre},
  J.~E. 2014, \apjl, 795, L5

\bibitem[{{The Astropy Collaboration} {et~al.}(2018){The Astropy
  Collaboration}, {Price-Whelan}, {Sip{\H o}cz}, {G{\"u}nther}, {Lim},
  {Crawford}, {Conseil}, {Shupe}, {Craig}, {Dencheva}, {Ginsburg},
  {VanderPlas}, {Bradley}, {P{\'e}rez-Su{\'a}rez}, {de Val-Borro}, {Aldcroft},
  {Cruz}, {Robitaille}, {Tollerud}, {Ardelean}, {Babej}, {Bachetti}, {Bakanov},
  {Bamford}, {Barentsen}, {Barmby}, {Baumbach}, {Berry}, {Biscani}, {Boquien},
  {Bostroem}, {Bouma}, {Brammer}, {Bray}, {Breytenbach}, {Buddelmeijer},
  {Burke}, {Calderone}, {Cano Rodr{\'{\i}}guez}, {Cara}, {Cardoso},
  {Cheedella}, {Copin}, {Crichton}, {D{\'A}vella}, {Deil}, {Depagne},
  {Dietrich}, {Donath}, {Droettboom}, {Earl}, {Erben}, {Fabbro}, {Ferreira},
  {Finethy}, {Fox}, {Garrison}, {Gibbons}, {Goldstein}, {Gommers}, {Greco},
  {Greenfield}, {Groener}, {Grollier}, {Hagen}, {Hirst}, {Homeier}, {Horton},
  {Hosseinzadeh}, {Hu}, {Hunkeler}, {Ivezi{\'c}}, {Jain}, {Jenness}, {Kanarek},
  {Kendrew}, {Kern}, {Kerzendorf}, {Khvalko}, {King}, {Kirkby}, {Kulkarni},
  {Kumar}, {Lee}, {Lenz}, {Littlefair}, {Ma}, {Macleod}, {Mastropietro},
  {McCully}, {Montagnac}, {Morris}, {Mueller}, {Mumford}, {Muna}, {Murphy},
  {Nelson}, {Nguyen}, {Ninan}, {N{\"o}the}, {Ogaz}, {Oh}, {Parejko}, {Parley},
  {Pascual}, {Patil}, {Patil}, {Plunkett}, {Prochaska}, {Rastogi}, {Reddy
  Janga}, {Sabater}, {Sakurikar}, {Seifert}, {Sherbert}, {Sherwood-Taylor},
  {Shih}, {Sick}, {Silbiger}, {Singanamalla}, {Singer}, {Sladen}, {Sooley},
  {Sornarajah}, {Streicher}, {Teuben}, {Thomas}, {Tremblay}, {Turner},
  {Terr{\'o}n}, {van Kerkwijk}, {de la Vega}, {Watkins}, {Weaver}, {Whitmore},
  {Woillez}, \& {Zabalza}}]{astropy}
{The Astropy Collaboration}, {Price-Whelan}, A.~M., {Sip{\H o}cz}, B.~M.,
  {et~al.} 2018, ArXiv e-prints, arXiv:1801.02634

\bibitem[{{Tikhonov} \& {Klypin}(2009)}]{Tikhonov09}
{Tikhonov}, A.~V., \& {Klypin}, A. 2009, \mnras, 395, 1915

\bibitem[{{Tollerud} {et~al.}(2015){Tollerud}, {Geha}, {Grcevich}, {Putman}, \&
  {Stern}}]{tollerud15}
{Tollerud}, E.~J., {Geha}, M.~C., {Grcevich}, J., {Putman}, M.~E., \& {Stern},
  D. 2015, \apjl, 798, L21

\bibitem[{{Tollerud} {et~al.}(2016){Tollerud}, {Geha}, {Grcevich}, {Putman},
  {Weisz}, \& {Dolphin}}]{tollerud16}
{Tollerud}, E.~J., {Geha}, M.~C., {Grcevich}, J., {et~al.} 2016, \apj, 827, 89

\bibitem[{{Tollerud} \& {Peek}(2018)}]{Tollerud18}
{Tollerud}, E.~J., \& {Peek}, J.~E.~G. 2018, \apj, 857, 45

\bibitem[{{Tolstoy}(1999)}]{Tolstoy99}
{Tolstoy}, E. 1999, in IAU Symposium, Vol. 192, The Stellar Content of Local
  Group Galaxies, ed. P.~{Whitelock} \& R.~{Cannon}, 218

\bibitem[{{Tonnesen} \& {Bryan}(2012)}]{Tonnesen12}
{Tonnesen}, S., \& {Bryan}, G.~L. 2012, \mnras, 422, 1609

\bibitem[{{Tully} {et~al.}(2013){Tully}, {Courtois}, {Dolphin}, {Fisher},
  {H{\'e}raudeau}, {Jacobs}, {Karachentsev}, {Makarov}, {Makarova},
  {Mitronova}, {Rizzi}, {Shaya}, {Sorce}, \& {Wu}}]{tully13}
{Tully}, R.~B., {Courtois}, H.~M., {Dolphin}, A.~E., {et~al.} 2013, \aj, 146,
  86

\bibitem[{{Wegner}(2000)}]{Wegner00}
{Wegner}, W. 2000, \mnras, 319, 771

\bibitem[{{Weisz} {et~al.}(2014){Weisz}, {Dolphin}, {Skillman}, {Holtzman},
  {Gilbert}, {Dalcanton}, \& {Williams}}]{Weisz14}
{Weisz}, D.~R., {Dolphin}, A.~E., {Skillman}, E.~D., {et~al.} 2014, \apj, 789,
  147

\bibitem[{{Weisz} {et~al.}(2011){Weisz}, {Dalcanton}, {Williams}, {Gilbert},
  {Skillman}, {Seth}, {Dolphin}, {McQuinn}, {Gogarten}, {Holtzman}, {Rosema},
  {Cole}, {Karachentsev}, \& {Zaritsky}}]{Weisz11}
{Weisz}, D.~R., {Dalcanton}, J.~J., {Williams}, B.~F., {et~al.} 2011, \apj,
  739, 5

\bibitem[{{Wetzel} {et~al.}(2015){Wetzel}, {Tollerud}, \& {Weisz}}]{Wetzel15}
{Wetzel}, A.~R., {Tollerud}, E.~J., \& {Weisz}, D.~R. 2015, \apjl, 808, L27

\bibitem[{{Wheeler} {et~al.}(2019){Wheeler}, {Hopkins}, {Pace},
  {Garrison-Kimmel}, {Boylan-Kolchin}, {Wetzel}, {Bullock}, {Kere{\v{s}}},
  {Faucher-Gigu{\`e}re}, \& {Quataert}}]{Wheeler19}
{Wheeler}, C., {Hopkins}, P.~F., {Pace}, A.~B., {et~al.} 2019, \mnras, 490,
  4447

\end{thebibliography}

\end{document}